# Monolithic Silicon-Photonics Linear-Algebra Accelerators Enabling Next-Gen Massive MIMO

Tzu-Chien Hsueh, *Senior Member*, *IEEE*, Yeshaiahu Fainman, *Fellow*, *IEEE*, and Bill Lin, *Senior Member*, *IEEE*

*Abstract*—A system-on-chip (SoC) photonic-electronic linear-algebra accelerator with the features of wavelength-division-multiplexing (WDM) based broadband photodetections and high-dimensional matrix-inversion operations fabricated in advanced monolithic silicon-photonics (M-SiPh) semiconductor process technology is proposed to achieve substantial leaps in computation density and energy efficiency, including realistic considerations of energy/area overhead due to electronic/photonic on-chip conversions, integrations, and calibrations through holistic co-design methodologies to support linear-detection based massive multiple-input multiple-output (MIMO) decoding technology requiring the inversion of channel matrices and other emergent applications limited by linear-algebra computation capacities.

## I. INTRODUCTION

THE newly available millimeter wave (mmWave) spectrum in emerging 5G/6G wireless systems, coupled with advanced multiuser massive multiple-input multiple-output (MIMO) technology will make possible new wireless applications with massive throughput and robust low latency requirements in support of many-user mobility, including new wireless extended reality (XR) experience for large group-based applications. Such advanced MIMO-based wireless XR platforms will enable many users in large groups to freely roam through common areas while jointly experiencing interactive, immersive virtual, or mixed reality environments. While such wireless XR experience has potentially many applications in training, education, and entertainment, delivering such high resolution digital XR experience requires tremendous data rates that necessitate aggressive use of multiple antennas so that the required data rates can be achieved by means of parallel data streams, but in turn, the dimension of matrices involved in the MIMO decoding process increases substantially. In particular, linear-detection based MIMO decoding approaches require the inversion of channel matrices that are estimated by the base-station at a timescale fast enough to accurately capture potentially rapidly changing channel conditions. Unfortunately, the matrix-inversion problem has cubic complexity in the worst-case with $M$, the number of users that need to be served simultaneously, making the performance of conventional digital electronics approaches a key limiting factor to future scaling of massive MIMO systems.

To realize matrix-inversions on a single chiplet with massive computation capacities as mentioned, photonic computing through monolithic silicon-photonics (M-SiPh) fabrication and integration [1], [2] is the primary integrated-circuit (IC) fabrication platform of the proposed linear-algebra accelerator based on the major evidence as follows: (I) Since demanding data bandwidth requirements and the maturity of photonic IC developments, optical technology has been broadly used for high-volume data communications [3], [4], [5], [6], [7], [8]. Also, due to the advancement of massive MIMO far outpacing the Moore's law [9] and energy/area limitations of classical von Neumann computing architectures, wavelength-division-multiplexing (WDM) [10] based optical communication systems with on-chip optical devices and circuits, owning inherent parallelism, high degree of connectivity, and speed-of-light propagation, have been broadly adopted in the computation tasks of linear-algebra calculations, passive Fourier transforms [11], and matrix operations [9], [12], [13], [14], [15], [16], [17], [18], [19], [49] which exhibit superior photonic computing performances in terms of bandwidth density, processing latency, silicon area, and power consumption. (II) The availability of commercial M-SiPh process technology, GlobalFoundries 45SPCLO [1], [2], offers an opportunity to explore holistic co-design methodologies leveraging complementary capabilities of CMOS electronics and photonics to break through the development of computing systems currently at a crossroad. Especially, the M-SiPh technology possesses the capability of integrating all advanced electronic and photonic devices/circuits required in the proposed linear-algebra accelerator on a single chiplet, which can tremendously minimize the data-conversion and heterogeneous-interface overhead due to I/O circuits, electrostatic-discharge (ESD) protection diodes, chip bumps/pads, interposers, packages, and bonding-wires among separate electronic and photonics dies. These inevitable downsides of the heterogeneous integrations mostly have been excluded in the performance metrics of photonic computing literatures [9], [13], [14], [15], [16], [17], [18], [19], but revealed by the limited computation scalabilities in their hardware demonstrations.

The goal of the proposed M-SiPh linear-algebra accelerators is to practically implement a well-interfaced and energy/area efficient SoC with the functionalities of high-dimensional matrix-vector multiplications (MVM), matrix-matrix multiplications (MMM), matrix-matrix additions (MMA), and eventually matrix-inversions (MI) for the wireless-channel estimations in the next-generation massive MIMO. The remainder of the paper is organized as follows. The background of advanced massive MIMO is summarized in Section II. The motivation of using M-SiPh technology and the MI approximation algorithm are elaborated in Section III. The architecture and building-block functionality with performance specifications of an M-SiPh MVM accelerator are analyzed in Section IV. The architecture scalability, parallelism, and realization of M-SiPh MMM and final M-SiPh MI accelerators are described in Section V and VI, respectively. The performance evaluation and conclusion are summarized in Section VII.

## II. BACKGROUND OF ADVANCED MASSIVE MIMO

Consider a large-scale uplink multiuser massive MIMO system with $N$ antennas at the base-station and $M$ ($< N$) single antenna users, each user transmitting a symbol from an $m$-QAM constellation set. The resulting transmit vector is denoted by $\pmb{X} = [x_1, x_2, \dots, x_M]^T$, and the received vector on the base-station side is denoted by $\pmb{U} = [u_1, u_2, \dots, u_N]^T$. The system model for the MIMO uplink can be expressed as follows:

$$\pmb{U} = \pmb{H} \cdot \pmb{X} + \pmb{NOISE} \tag{1}$$

where $\pmb{H}$ is an $N \times M$ complex-value channel matrix, and $\pmb{NOISE}$ is an additive noise vector. The entries in both $\pmb{H}$ and $\pmb{NOISE}$ are typically assumed to be independent and identically distributed (i.i.d.) zero-mean unit-variance complex Gaussian random variables.

In order to compute the soft-estimates in the form of logarithm likelihood ratios (LLRs) for the coded bit streams, given $\pmb{H}$ and $\pmb{U}$, the linear detection method is often employed. This algorithm first constructs an $M \times M$ Gram matrix $\pmb{Z} = \pmb{H}^H \cdot \pmb{H}$, so that the linear detection estimation of the transmitted vector can be computed as follows:

$$\hat{\pmb{X}} = \pmb{Z}^{-1} \cdot \pmb{H}^H \cdot \pmb{U} \tag{2}$$

## III. CO-DESIGN OF MATRIX-INVERSIONS WITH M-SIPH LINEAR-ALGEBRA ACCELERATORS

Although multiple prior works have developed photonic accelerators in the field of neuromorphic computing for convolutional neural networks (CNN) and recurrent neural networks (RNN) [13], [14], [15], [16], [17], [18], [19], many challenges have to be addressed when utilizing photonic accelerators in the applications of massive MIMO channel estimations requiring high-dimensional MMM computations: (I) Prior work on photonic accelerators for CNN and RNN mainly focused on MVM computations with *static* weights whose matrix elements are rarely reprogrammed once the training process has completed. However, the repetition process in a matrix-inversion approximation algorithm (e.g., Neumann-series approximation) is different: the degree of the approximation accuracy can be enhanced by the repetition number of the algorithm execution *dynamically* taking the previous repetition result as the input of current repetition. (II) The MI calculations involve not only a high-dimensional MMM but also MMA. If an MI accelerator is implemented in a naïve way with photonic MMM and MMA independently, the linear projection of inputs and the calculation of the scaled dot-products would require a certain amount of data conversions between the electronic and photonic domains back and forth. (III) Most of prior work on photonic accelerators assumed separate chip implementations for the photonic and electronic parts with heterogeneous integrations which could incur expensive chip-to-chip communication hardware and degrade the effectiveness of a photonic computing approach to the linear-algebra acceleration as mentioned in Section I.

The novelties and methodologies elaborated in Section IV, V, and VI aim to address these significant challenges by performing MVM, MMM and MMA purely in the photonic domain on a single M-SiPh chiplet without intermediate optical memory and optical-to-electrical-to-optical (O/E/O) conversions to demonstrate the feasibility of the M-SiPh linear-algebra accelerator for supporting next-generation massive MIMO workloads with two-layer enhancements of energy, area, and computation throughput: (I) The cost reduction of end-to-end computations across electronic and photonics domains through the M-SiPh integration. (II) The energy/area reduction through the hardware-friendly MI approximation algorithm elaborated in the rest of this section.

As shown in Eq. (2), the key computational bottleneck in the MIMO linear detection process is computing the inverse of $\pmb{Z}$. In particular, the computation of $\pmb{Z}^{-1}$ using exact inversion methods, such as Cholesky decomposition [21], [22], requires $O(M^3)$ operations, which is very expensive to realize in hardware when an increasingly large number of users need to be served simultaneously (i.e., with increasing $M$). Further, in mmWave settings, channel conditions can change very rapidly, making the need to perform the inversion of large matrices at increasingly smaller timescales.

To scale to large values of $M$ (e.g., $M \geq 32$) for massive MIMO, this paper proposes a combination of photonic computing to perform efficient MI calculations and algorithmic-efficient approximation based on the Neumann series [23] to obtain the MI result with required accuracy. The Neumann-series approximation approach exploits the property that $\pmb{Z}$ is an almost diagonal matrix in massive MIMO systems. In particular, $\pmb{Z}$ can be decomposed as $\pmb{Z} = \pmb{D} + \pmb{E}$, where $\pmb{D}$ is a diagonal matrix with the diagonal entries of $\pmb{Z}$, and $\pmb{E}$ is the

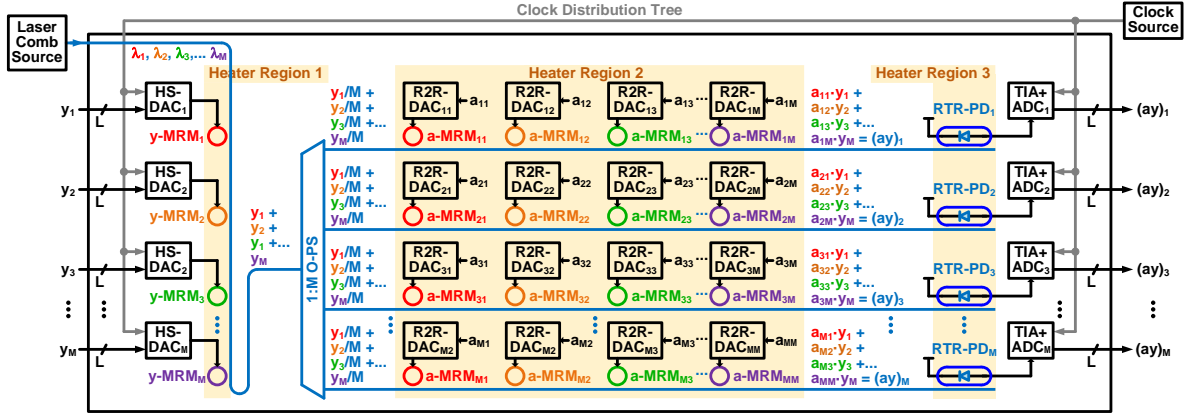

Fig. 1. The system block diagram of the M-SiPh MVM accelerator.

off-diagonal part of $\mathbf{Z}$. Then, the Neumann series to compute the inverse can be expressed as:

$$\widetilde{\mathbf{Z}}_k^{-1} = \sum_{n=0}^{k-1} (-\mathbf{D}^{-1} \cdot \mathbf{E})^n \cdot \mathbf{D}^{-1} \qquad (3)$$

where $k$ is the number of repetitions to be computed in the series and $\widetilde{\mathbf{Z}}_k^{-1}$ is the $k$-term approximation of $\mathbf{Z}^{-1}$. Let $\mathbf{Y}[k] = \widetilde{\mathbf{Z}}_k^{-1}$, $\mathbf{A} = -\mathbf{D}^{-1}\mathbf{E}$, and $\mathbf{B} = \mathbf{D}^{-1}$. Then, Eq. (3) can be rewritten with the following recurrence equation:

$$\mathbf{Y}[k] = \mathbf{B} + \mathbf{A} \cdot \mathbf{Y}[k-1], \qquad \forall\, k \geq 1 \qquad (4)$$

with $\mathbf{Y}[0] = \mathbf{0}$. This recurrence computation can be iteratively computed via repeated MMM and MMA operations.

In particular, this paper proposes to efficiently implement Eq. (4) in M-SiPh for the Neumann-series approximation, specifically the MMM and MMA, i.e., "$\mathbf{A} \cdot \mathbf{Y}$" and "$\mathbf{B} +$" in Eq. (3), respectively, in the photonic domain at the speed of light. Besides accelerating the Neumann-series approximation in M-SiPh process technology, the future work will also explore other MIMO detection algorithms by means of M-SiPh linear-algebra acceleration, including the successive-over-relaxation [24], Gauss-Seidel [25], optimized coordinated descent [26], conjugate-gradient [27], Richardson [28], and Jacobi [29] methods.

## IV. M-SiPh Matrix-Vector Multiplications

A unified M-SiPh MVM accelerator serves as the primary functional block of the M-SiPh MMM, MMA, and eventually M-SiPh MI in the MIMO channel estimation with high degree of reconfigurability in terms of the internal matrix weights and on-chip interconnections. The M-SiPh MVM functionality is basically realized by utilizing the high-degree spatial parallelism of light-waves and the concept of optical WDM technique [10] as illustrated in Fig. 1. A laser comb source injects "$M$" wavelengths through a single waveguide into a M-SiPh MVM accelerator consisting of "$M$" L-bit high-speed digital-to-analog data converters (HS-DAC), "$M^2$" low-power static DACs (R2R-DAC), "$M$" transimpedance amplifiers (TIA) individually followed by "$M$" L-bit analog-to-digital data converters (ADC), digital registers, clock distribution, and discrete-time iteration mechanism in the electronic domain as well as "$M$" vector micro-ring modulators (y-MRM) for the input vector E/O conversion, "$M^2$" matrix micro-ring modulators (a-MRM) for the matrix E/O conversion and WDM-based MVM operations, "$M$" racetrack-resonator photodetectors (RTR-PD), 1-to-$M$ optical power splitter (O-PS), and waveguides (blue lines) in the photonic domain.

The $M$-by-1 input vector is denoted by $\mathbf{Y}_{M\times 1}$ with its elements $y_i$, $i = 1 \ldots M$; the $M$-by-1 output vector is denoted by $(\mathbf{AY})_{M\times 1}$ with its elements $(ay)_i$, $i = 1 \ldots M$; the $M$-by-$M$ primary matrix is denoted by $\mathbf{A}_{M\times M}$ with its elements $a_{ij}$, $i$ and $j = 1 \ldots M$ respectively. Mathematically, this MVM functionality is expressed as $\mathbf{A}_{M\times M} \cdot \mathbf{Y}_{M\times 1} = (\mathbf{AY})_{M\times 1}$. The operational timings of input and output vectors are both managed by the electronic circuits in the digital domain for the seamless compatibility and interface-relation between the M-SiPh accelerator and other on-chip digital application-specific integrated-circuits (ASIC), processors, lookup tables, and memory. The circuit and interconnection details of the 1st MVM row from the input $y_1$ to output $(ay)_1$ within the M-SiPh MVM accelerator is shown in Fig. 2 and further elaborated in the following sub-sections. Note that the clock period per MVM operation, setting the computation speed and throughput, is determined by the signal propagation latency from the clock edge launching the computation data input from each HS-DAC through the MRM, O-PS, WDM-based MVM operation, RTR-PD, TIA, to the ADC data output sampled by the next clock edge. Therefore, the bandwidths of all electronic and photonic circuits/devices within this critical signal path aggregately decide the limitations of the clock period and eventually computation throughput (MAC/s) of this M-SiPh MVM accelerator, which is dominated by the electronic circuits analyzed in the following sub-sections.

### A. Optical WDM-Based MVM Architecture

The M-SiPh MVM functionality is established on the theory and concept of optical WDM incoherent data transmission technique [3], [6], [10]. The data carries are "$M$" of laser light-waves with identical optical power but individual wavelengths $\lambda_i$, $i = 1 \ldots M$, with pre-defined spectrum spacing between adjacent wavelengths to assure the signal integrity of incoherent data transmission. As shown in the left of Fig. 1, the powers of these incoming light-waves propagating through a single waveguide are individually and sequentially modulated by the vector MRMs y-MRM$_i$ controlled by HS-DAC$_i$, $i = 1 \ldots M$, so that each $y_i$ of the input data vector $\boldsymbol{Y}_{M\times 1}$ from the electronic domain is correspondingly converted into the light-wave power of $\lambda_i$ linearly proportional to $y_i$ with a consistent scalar across all $i = 1 \ldots M$, which are the E/O conversion processes. At the input of the O-PS in Fig. 1, the aggregate light-wave power is proportional to $\sum_{i=1}^{M} y_i = (y_1 + y_2 + y_3 + \ldots + y_M)$. Since the O-PS evenly splits this aggregate light-wave power into its "$M$" fan-outs, the light-wave powers in the waveguides of all MVM rows shall be identical and proportional to $\sum_{j=1}^{M} y_j/M = (y_1 + y_2 + y_3 + \ldots + y_M)/M$. Note that the wavelength index for the "i-th" MVM row is "j" not "i".

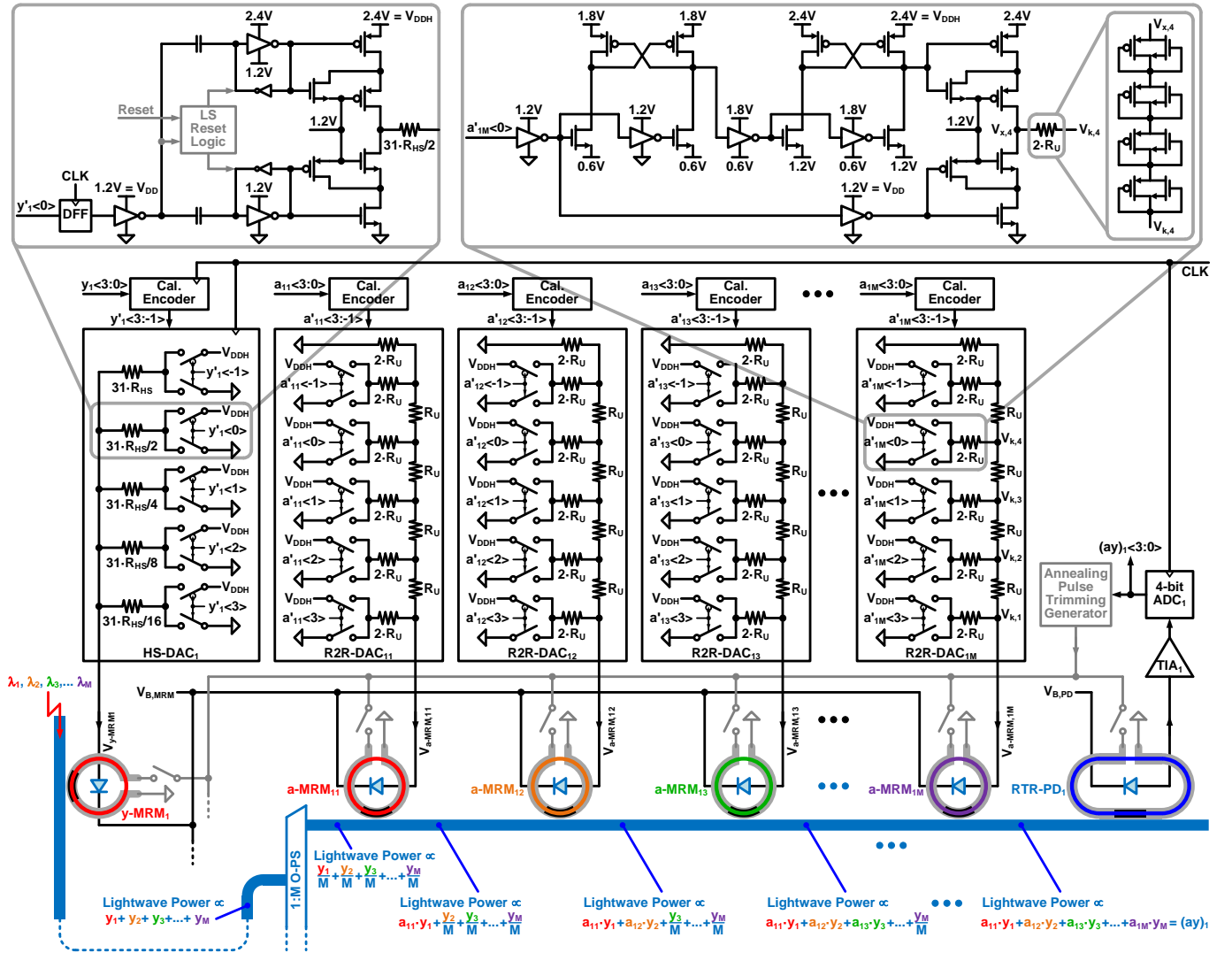

Fig. 2. The zoom-in version of the 1st M-SiPh MVM row from the input $y_1<3:0>$ to output $(ay)_1<3:0>$ with detailed circuits, interconnections, and clocking relation among the HS-DAC, R2R-DACs, MRMs, RTR-PD, and post-fabrication trimming mechanism.

After the O-PS, the light-wave powers of all $\lambda_j$, $j = 1 \ldots M$, in the i-th MVM row will individually and sequentially go through the power modulation effects of "$M$" MRMs (i.e., a-MRM$_{ij}$, $j = 1 \ldots M$). Again, each a-MRM$_{ij}$ driven by $a_{ij}$ can only modulate the $\lambda_j$ light-wave power ($\propto y_j/M$) when the indexes $j$ of $a_{ij}$ and $y_j$ are matched. For instance, as shown in the bottom of Fig. 2 for the 1st row (i = 1) of the MVM operation, all $\lambda_j$ light-wave powers ($\propto \sum_{j=1}^{M} y_j/M$) together pass though the power modulation effect of a-MRM$_{11}$ (j = 1), but only $\lambda_1$ light-wave power ($\propto y_1/M$) gets modulated to be proportional to $(a_{11}\cdot y_1)$. The rest of $\lambda_j$, $j = 2 \ldots M$, light-wave powers can be all-pass filtered through a-MRM$_{11}$ without any power change. After passing through all a-MRM$_{1j}$, $j = 1 \ldots M$, all $\lambda_j$ light-wave powers in the 1st MVM row can be respectively modulated based on $a_{1j}$, $j = 1 \ldots M$, at the speed of light. At the input port of RTR-PD$_1$, the aggregate power on the 1st waveguide becomes proportional to $\sum_{j=1}^{M} a_{1j}\cdot y_j = (a_{11}\cdot y_1 + a_{12}\cdot y_2 + a_{13}\cdot y_3 + \ldots + a_{1M}\cdot y_M) = (ay)_1$, which can represent the equivalent dot-product of the input vector $\boldsymbol{Y}_{M\times 1}$ and the 1st row vector of matrix $\boldsymbol{A}_{M\times M}$. By replicating the same process across all row vectors of $\boldsymbol{A}_{M\times M}$ in parallel, the total light-wave power of each MVM row can be eventually proportional to $(ay)_i$ with a consistent scalar across all $i = 1 \ldots M$. After the following O/E and A/D conversions through RTR-PD$_i$ and ADC$_i$, respectively, the MVM operation is essentially completed and all elements (ay)$_i$, i = 1 … M of output vector $(AY)_{M \times 1}$ is sampled and digitally preserved in the electronic domain.

Both WDM-based communication and computation systems rely on multi-wavelengths simultaneously carrying data information through a communication channel to maximize data communication and computation capacity, respectively. Meanwhile, they have two major differences: (I) WDM-based communication modulates each light-wave power once for the purpose of E/O conversion, but WDM-based computation needs to modulate each light-wave power at least twice to perform the E/O conversions and then equivalent operation of multiplications. (II) The light-wave power detections in WDM-based communication are designated to distinguish individual light-wave powers for recovering the data information carried by each light-wave. Although WDM-based computation only needs to detect the aggregate light-wave power instead of individuals, the wavelength isolations across the WDM spectrum still have to be well maintained since the multiplication operations are done by independent light-wave power modulations in the presence of all light-waves. In addition, the equivalent summation of the dot-products requires consistent power-abortion photodetections across the entire WDM spectrum to maintain the computation linearity during the O/E conversions. These two major differences between WDM-based communication and computation determine the E/O/E modulation/detection speeds and D/A/D data resolutions of these two systems.

### B. High-Speed Digital-to-Analog Converters

In the high-speed computation path, each element y$_i$ of the input vector $Y_{M \times 1}$ denoted by L-bit digital data is firstly converted into an analog-voltage level through HS-DAC$_i$ to drive y-MRM$_i$ for λ$_i$ light-wave power modulation as the E/O conversion process. Within the maximized the E/O dynamic range (DR), the issue of nonlinear MRM transmission power-gain across all possible $2^L$ voltage-levels can be alleviated by adding one calibration bit in HS-DAC$_i$ with a digital calibration encoder to map the incoming L-bit y$_i$ data to (L+1)-bit y'$_i$ data for better E/O conversion linearity while maintaining the same amount of $2^L$ voltage-levels, not $2^{L+1}$, at the HS-DAC$_i$ output. This E/O linearity improvement technique is further discussed in Section IV.D.

Every HS-DAC is realized by a large-swing voltage-mode source-series terminated (SST) driver architecture [30] to reach up to GHz sampling rates, low static-power consumption, and rail-to-rail voltage driving capability for maximizing the light-wave power modulation DR of each corresponding MRM. As shown in the left of Fig. 2, the binary-weighted driver-segments driven by y'$_1$<3:-1> in HS-DAC$_1$ have identical architecture but are reciprocally scaled according to their own series-resistors. For instance, the driver-segment of y'$_1$<3> is formed by sixteen driver-segments of y'$_1$<-1>. According to the binary data of y'$_1$<3:-1>, some driver-segments short their series-resistors to V$_{DDH}$ through the cascode push-up PMOS transistors, and the rest to GND through the cascode pull-down NMOS transistors. Equivalently, all the parallel push-pull resistances between V$_{DDH}$ and GND together form a voltage-divider at the HS-DAC$_1$ output, which controls the reverse biased voltage (V$_{B,MRM}$ − V$_{y\text{-}MRM1}$) of the y-MRM$_1$ P/N junction used to modulate the λ$_1$ transmission power. Note that AC-coupled level-shifters are required for push-pull pre-drivers as shown in the top-left of Fig. 2 [3] to enable high-speed single-stage 2× voltage level-shifting from V$_{DD}$ regular digital supply to V$_{DDH}$ high-voltage supply. To maximize the E/O DR up to V$_{DDH}$ and satisfy 45-nm CMOS reliability requirements in 45SPCLO, the AC-coupled level-shifters and cascode push-pull driver architecture are necessary to maintain all transistors operating within the 1.2-V terminal-to-terminal voltages.

The speed and bandwidth specifications of HS-DAC$_i$ are based on two aspects: (I) The circuit latency from the clock edge of the y'$_i$<3:-1> register (or DFF) to the cascode push-pull drivers, including the delays of DFF clock-to-output, digital buffers, AC-coupled level-shifters, and push-pull pre-drivers, is about 100 ps in 45SPCLO. This circuit latency of HS-DAC$_i$ can be 1$^{st}$-orderly cancelled by delaying the sampling clock edge of the ADC$_i$ accordingly. (II) The RC time-constant of the HS-DAC$_i$ output network is determined by the driver AC resistance = (31·R$_{HS}$/16 ∥ 31·R$_{HS}$/8 ∥ 31·R$_{HS}$/4 ∥ 31·R$_{HS}$/2 ∥ 31·R$_{HS}$) = R$_{HS}$ and equivalent lump capacitance contributed by the transistors, resistors, and y-MRM$_i$, C$_{HS\text{-}DAC}$ + C$_{MRM}$ ≈ 30 fF [7]. For 2-GSym/s, 4-bit, and 2.4-V DR data, if HS-DAC$_i$ should spend < 200 ps settling its output to a static voltage level within a half least-significant-bit (LSB/2) of the 4-bit DR, i.e., exp(−200-ps/τ) < 1/($2^{4+1}$). Then, the time-constant of the HS-DAC$_i$ output network, τ = R$_{HS}$·30-fF, needs to be < 58 ps, and thus R$_{HS}$ should be < 58-ps/30-fF ≈ 2 kΩ. In other words, if this D/A conversion time is designed to spend < 200 ps out of the 500-ps clock-period budget, this bandwidth specification leads to the result of R$_{HS}$ < 2 kΩ and corresponding static power consumption discussed below.

Both dynamic and static power consumptions of each HS-DAC shall be considered: (I) The digital logics, including the DFFs, calibration encoder, data buffers, and push-pull pre-drivers, consume around 0.2-mW dynamic power at 2-GSamp/s following the convention of C$_{load}$·V$^2_{supply}$·f$_{clock}$/2 with the assumption of 50% logical-HIGH and 50% logical-LOW data pattern per digital gate during the regular MVM operations. (II) The static power consumption is mainly due to the DC current path from V$_{DDH}$ to GND of the voltage-divider formed by the parallel push-pull P/N transistors and series-resistors as mentioned. Note that although the driver AC resistance R$_{HS}$ stays constant, the DC path resistance from V$_{DDH}$ to GND is data-dependent. Again, if the data patterns of y$_i$ with corresponding HS-DAC$_i$ output voltage-levels are uniformly occurred between GND to V$_{DDH}$ during the regular MVM operations, the average static power consumption per HS-DAC can be derived as follows:

$$P_{HS\text{-}DAC,ST} = \frac{V_{DDH}^2}{R_{HS}} \cdot \frac{\sum_{k=1}^{2^L-1}(k-1) \cdot (2^L - k)}{2^{L-1} \cdot (2^L - 1)^2} \quad (5)$$

If L = 4 bits, $R_{HS} \approx 2$ kΩ, and $V_{DDH}$ = 2.4 V as shown in Fig. 2, each HS-DAC would consume 0.45-mW static power and total 0.65 mW in average when the dynamic power is also included.

Including the cascode voltage-mode drivers, push-pull pre-drivers, unsilicided poly-resistors, calibration encoder, and digital logics for resetting the initial conditions of the AC-coupled level-shifters [3], the silicon area of HS-DAC$_i$ is around 50-um×20-um.

*C. Low-Power R2R Digital-to-Analog Converters*

The multiplication of each input vector-element and matrix-element $a_{ij} \cdot y_j$ in the photonic domain is realized by the secondary light-wave power modulation to provide a transmission power-gain proportional to $a_{ij}$ on the top of the $\lambda_j$ light-wave whose power has been pre-modulated and split to be proportional to $y_j/M$ in the i-th waveguide. In other words, after $y_i$ is E/O converted through HS-DAC$_i$ and y-MRM$_i$ and then evenly power split into the i-th MVM row, the element-to-element multiplication is done by another light-wave power modulation through R2R-DAC$_{ij}$ and a-MRM$_{ij}$ only effective to the $\lambda_j$ light-wave power. Apparently, R2R-DAC$_{ij}$ is used to convert the digital multiplicand $a_{ij}$ to its corresponding voltage-level for setting the transmission power-gain of a-MRM$_{ij}$, which is the same D/A and E/O operations in Section IV.B. Note that the bandwidth of R2R-DAC$_{ij}$ for converting $a_{ij}$ is not critical since the value of the matrix $\boldsymbol{A}_{M \times M}$ has been pre-determined and stays static during the regular MVM operations as described in Section III for the Neumann-series approximation. This fact beneficially allows to use low-power small-area R-2R voltage-divider architecture [31] to implement $M^2$ of DACs with $M^2$ of MRMs on a single chiplet for such a high-dimensional M-SiPh MVM accelerator as shown in Fig. 1.

Similar to the linearization technique used in HS-DAC$_i$, the nonlinear transmission power-gain of a-MRM$_{ij}$ across all $2^L$ voltage-levels is alleviated by adding one calibration bit, so R2R-DAC$_{ij}$ with a digital calibration encoder can map the L-bit $a_{ij}$ to (L+1)-bit a'$_{ij}$ to enhance the E/O conversion linearity and maintain the same amount of $2^L$ voltage-levels, not $2^{L+1}$. R2R-DAC$_{ij}$ shown in Fig. 2 is built by alternating series-R and shunt-2R resistances with push-pull cascode drivers to perform a voltage-divider driving the capacitive electrode of a-MRM$_{ij}$. This R-2R voltage-divider architecture is very beneficial to energy/area efficiency by taking advantage of the static operation, low-bandwidth requirement, and minimal amounts of resistors/transistors to generate all required voltage-levels.

The circuit implementation of each shunt-2R segment in R2R-DAC$_{ij}$ is similar to the driver segment in HS-DAC$_i$, which utilizes a large-swing voltage-mode driver architecture with push-pull cascode transistors to short the $2 \cdot R_U$ resistor in each segment to $V_{DDH}$ or GND. By taking advantage of the low-bandwidth operation, the push-up PMOS transistor can be driven by a two-stage static level-shifter as shown in the top-right of Fig. 2 to accommodate the E/O DR up to 2.4 V and maintain 45-nm CMOS reliability in 45SPCLO with negligible power consumption. The primary power consumption is the static current in the voltage-dividers formed by the R-2R network according to the digital multiplicand $a_{ij}$. To express the average static power consumption of R2R-DAC$_{ij}$, multiple indexes and variables are pre-defined as follows: "i" and "j" are row and column indexes of the matrix, respectively, but do not involve in the power calculation; k = 1 … $2^L$ is the voltage-level index; p = 1 … L is the circuit-node index, q = 1 … L is the Kirchhoff's Voltage Law (KVL) superposition index of the "p-th" circuit-node; $R_U$ is the R-2R unit-resistor as shown in Fig. 2; $R_p = (G_p/H_p) \cdot R_U$ is the one-side equivalent resistance of the "p-th" circuit-node; $G_p$ and $H_p$ are integers; $G_p/H_p$ forms a simplest fraction; $V_{k,p}$ is the KVL superposition voltage of the "p-th" circuit-node with its "k-th" voltage-level. If the digital values of all $a_{ij}$ and their own R2R-DAC$_{ij}$ output voltage-levels are uniformly distributed between GND to $V_{DDH}$ across the entire matrix $\boldsymbol{A}_{M \times M}$, the average static power per R2R-DAC can be derived as follows:

$$G_1 = 1, H_1 = 0 \Rightarrow R_1 = \infty, \qquad p = 1$$
$$R_p = R_{p-1} || (2R_U) + R_U = \frac{G_p}{H_p} \cdot R_U, \quad p = 2 \ldots L$$
$$\frac{V_{k,p}}{V_{DDH}} = \sum_{q=1}^{L} a_{ij,k}\langle L-q \rangle \cdot \frac{\min[G_p, G_q]}{2^{p+q-1}}, \quad \begin{array}{l} k = 1 \ldots 2^L \\ p = 1 \ldots L \end{array}$$
$$P_{R2R-DAC,ST} = \frac{V_{DDH}^2}{R_U} \cdot \frac{\sum_{k=1}^{2^L}\sum_{p=1}^{L} a_{ij,k}\langle L-p\rangle \cdot \left(1 - \frac{V_{k,p}}{V_{DDH}}\right)}{2^{L+1}}$$
(6)

If L = 4 bits, $R_U \approx 5$ MΩ, $V_{DDH}$ = 2.4 V, R2R-DAC$_{ij}$ consumes about 7.2 µW. For a small silicon area and large resistance, each $2 \cdot R_U$ is implemented by four stacked P/N parallel diode-connected transistors operated in sub-threshold-region transistors [32] as active-resistor templates shown in the top-right of Fig. 2. Although the temperature coefficients and process-corner variations of these active resistors are relatively high, their resistance ratios in the R-2R network are actually tolerable within the 4-bit accuracy requirement. However, the nonlinearity due to data-dependent terminal voltages across these active resistors require extra attention. For instance, the $2 \cdot R_U$ between $V_{x,4}$ and $V_{k,4}$ in Fig. 2 is data-dependent because $V_{x,4}$ can be short to either $V_{DDH}$ or GND so $V_{k,4}$ can vary from GND to (85/128)$\cdot V_{DDH}$ according to the $V_{k,p}$ formula in Eq. (6). This issue can alleviated by the complimentary sub-threshold-region P/N active-resistor architecture proposed in [49]: (I) If $V_{x,4} = V_{DDH} > V_{k,4}$, the sub-threshold biases are $0 < |V_{GS,PMOS}| < V_{th,PMOS}$ and $V_{GS,NMOS} = 0$. (II) If $V_{x,4} =$ GND $< V_{k,4}$, the sub-threshold biases are $0 < |V_{GS,NMOS}| < V_{th,NMOS}$ and $V_{GS,PMOS} = 0$. In either case, the sub-threshold biases of the P/N transistor are always complimentary to cancel the first-order nonlinearity due to the data-dependent active resistance. In addition, the nonlinearity due to $2^L$ possible $V_{k,p}$ values based on $a_{ij}$<3:0> can be calibrated by the additional bit in a'$_{ij}$<3:-1> together with the a-MRM$_{ij}$ nonlinear transmission power-gain calibration discussed in Section IV.D.

Including the cascode voltage-mode drivers, push-up level-shifters, sub-threshold-region active resistors, and calibration encoder, the silicon area of R2R-DAC$_{ij}$ is about 20-µm×10-µm.

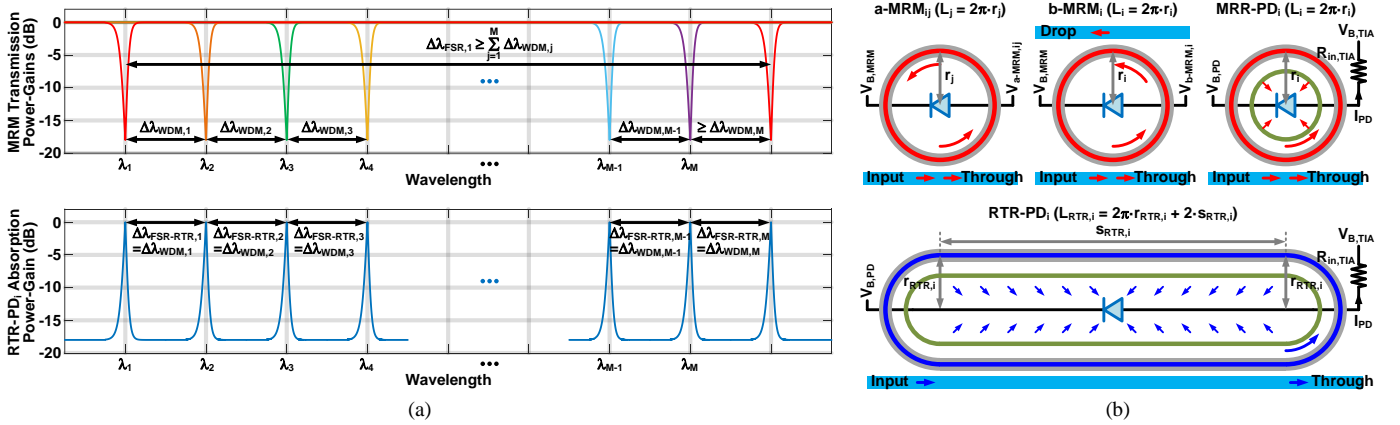

Fig. 3. (a) The transmission responses of a-MRM$_{ij}$, j = 1 ... M (or y-MRM$_i$, i = 1 ... M, when $\lambda_j$ is replaced with $\lambda_i$), and the absorption responses of RTR-PD$_i$. (b) The conceptual geometries and configurations of MRMs, MRR-PD and RTR-PD.

TABLE I
DESIGN TRADEOFF EXAMPLES AMONG MVM DIMENSIONS, WDM CROSSTALK, AND FABRICATION ERRORS

| M | $n_{eff}(\lambda_j)$ j = 1 ... M | $n_g(\lambda_j)$ j = 1 ... M | $m_{RTR,i}$ j = 1 ... M | $L_{RTR,i}$ | $\lambda_j$ j = 1 ... M | $\Delta\lambda_{WDM,j}$ j = 1 ... M | $m_j$ j = 1 ... M | $r_j$ j = 1 ... M | $\Delta r_{rms}/r_j$ Error Scale |
|---|---|---|---|---|---|---|---|---|---|
| 32 | 3.74 ~ 3.73 | 5.02 ~ 4.98 | 2321 ~ 2290 | 951.32 μm | 1534.5 ~ 1550 nm | 0.49 ~ 0.51 nm | 71 or 72 | 4.63 ~ 4.76 μm | 1× |
| 32 | 3.76 ~ 3.73 | 5.06 ~ 4.98 | 1160 ~ 1129 | 469.01 μm | 1519.0 ~ 1550 nm | 0.97 ~ 1.03 nm | 35 or 36 | 2.25 ~ 2.38 μm | 2× |
| 16 | 3.74 ~ 3.73 | 5.02 ~ 4.98 | 1160 ~ 1145 | 475.66 μm | 1535.0 ~ 1550 nm | 0.99 ~ 1.01 nm | 71 or 72 | 4.63 ~ 4.76 μm | 1× |

### D. Micro-Ring Modulators & E/O Conversion Linearity

Once an input vector-element or matrix-element ($y_i$ or $a_{ij}$) is converted to a voltage-level through HS-DAC$_i$ or R2R-DAC$_{ij}$ at the P-type electrode of their corresponding y-MRM$_i$ or a-MRM$_{ij}$, the power of the $\lambda_i$ or $\lambda_j$ light-wave located within the MRM resonance bandwidth can be effectively modulated based on the voltage delta (i.e., the reverse bias of the MRM P/N junction) between the N-type (i.e., $V_{B,MRM}$ in Fig. 2) and P-type (i.e., HS-DAC$_i$ or R2R-DAC$_{ij}$ output voltage) electrodes of the y-MRM$_i$ or a-MRM$_{ij}$, respectively.

The radius $r_{ij}$ design of a-MRM$_{ij}$ (or $r_i$ of y-MRM$_i$) in the whole WDM spectrum needs to satisfy at two fundamental requirements: (I) The minimum free spectral range (FSR) is determined by the number of the WDM wavelengths M and all WDM isolation spacing $\Delta\lambda_{WDM,j}$ between adjacent $\lambda_j$ [33] as shown in the top of Fig. 3(a). (II) The a-MRM$_{ij}$ resonance wavelength $\lambda_j$ under a particular resonance mode $m_j$ (an integer) is based on its effective refractive index $n_{eff}(\lambda_j)$, silicon propagation constant $\beta(\lambda_j)$, and ring circumference $L_j = 2\pi \cdot r_j$; i.e., if a resonance condition is satisfied for the MRM cavity $L_j$, a constructive interference is established by a certain wavelength having its round-trip phase shift equal to an integer multiple of $2\pi$ [34]. These two fundamental requirements are expressed in Eq. (7) and (8) as follows:

$$\Delta\lambda_{FSR,j} = \frac{\lambda_j^2}{n_g(\lambda_j) \cdot 2\pi \cdot r_j} \geq \sum_{j=1}^{M} \Delta\lambda_{WDM,j}$$
$$\Rightarrow r_j \leq \frac{\lambda_j^2}{n_g(\lambda_j) \cdot 2\pi \cdot \sum_{j=1}^{M} \Delta\lambda_{WDM,j}} \quad (7)$$

$$2\pi \cdot m_j = \beta(\lambda_j) \cdot L_j = \frac{2\pi}{\lambda_j} \cdot n_{eff}(\lambda_j) \cdot 2\pi \cdot r_j$$

$$\Rightarrow r_j = \frac{\lambda_j}{2\pi \cdot n_{eff}(\lambda_j)} \cdot m_j \quad (8)$$

$\Delta\lambda_{FSR,j}$ is the FSR of a-MRM$_{ij}$; $n_g(\lambda_j)$ is the silicon group index; $2\pi/\lambda_j$ is the free-space propagation constant; $\lambda_j$ presents the free-space resonance wavelength of a-MRM$_{ij}$ though the light-wave propagates in the silicon. An example for determining the radius $r_j$ of a-MRM$_j$ is elaborated in the rest of this sub-section as the initial design point without considering any post-process trimming or thermal control.

Since the wavelength-spectrum bandwidth in the silicon of 45SPCLO is in the range from 1180 nm to 1550 nm [6], and the MRM quality-factor Q is up to 8,000 [6], the WDM isolation spacing $\Delta\lambda_{WDM,j}$ across M = 32 wavelengths can be initially set to 0.5 nm, and the maximum WDM wavelength $\lambda_{32}$ is set to 1550 nm in this example. Note that the exact 32 resonance wavelengths $\lambda_j$, j = 1 … 32, distributed from 1534.5 nm to 1550 nm and 32 isolation spacing $\Delta\lambda_{WDM,j}$, j = 1 … 32, slightly varying from 0.49 nm to 0.51 nm listed in Table I are determined by the design equation of the broadband RTR-PD discussed in Section IV.E actually.

Therefore, based on all designated $\lambda_j$, $\Delta\lambda_{WDM,j}$, $n_{eff}(\lambda_j)$, and $n_g(\lambda_j)$ from the RTR-PD specification, the FSRs of all a-MRM$_{ij}$ must satisfy the requirement of $\Delta\lambda_{FSR,j} \geq 32 \cdot 0.5$ nm = 16 nm, so the upper bound of $r_j$ can be determined according to Eq. (7). Then, by picking proper mode integer $m_j$, $r_j$ of a-MRM$_{ij}$, j = 1 … M, can be obtained based on Eq. (8), which are in the range from 4.63 μm to 4.76 μm in this example. By setting different values of "M" and targeted $\Delta\lambda_{WDM,j}$, the design tradeoffs among the MVM dimension, WDM crosstalk, and MRM fabrication error are summarized in Table I including three scenarios. The 1st case has the worst WDM crosstalk due to the smallest $\Delta\lambda_{WDM,j}$. The 2nd case has the worst MRM fabrication error ∝ $(2\pi \cdot \Delta r_{rms})/(2\pi \cdot r_j)$ where $2\pi \cdot \Delta r_{rms}$ is the circumference error of a-MRM$_{ij}$ induced by random process variation. The 3$^{rd}$ case has the worst computation throughput because of the smallest value of "$M$" out of these three scenarios. Overall, including the central ring area and peripheral keep-out halo, the silicon area of a-MRM$_{ij}$ or y-MRM$_i$ can be within a 20-μm×20-μm tile.

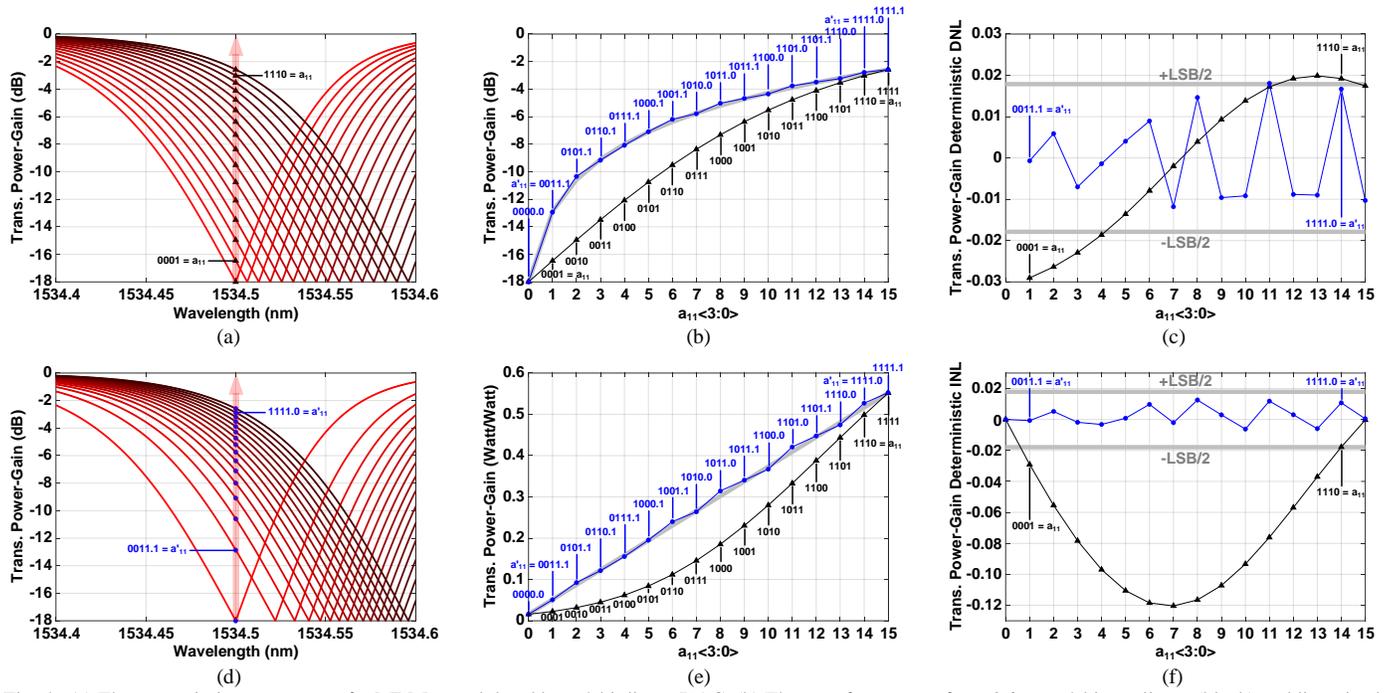

Fig. 4. (a) The transmission responses of a-MRM$_{11}$ modulated by a 4-bit linear DAC. (b) The transfer curves of a$_{11}$<3:0> vs. 4-bit nonlinear (black) and linearized (blue) E/O conversions on a dB scale. (c) The deterministic DNL of the 4-bit nonlinear (black) and linearized (blue) E/O conversions. (d) The transmission responses of a-MRM$_{11}$ modulated by a 4-bit nonlinear DAC for the linearization. (e) The identical transfer curves shown in (b) but on a linear scale. (f) The deterministic INL of the 4-bit nonlinear (black) and linearized (blue) E/O conversions.

The linearity issue of the E/O conversion process in the M-SiPh accelerator is extremely crucial. In particular, the deterministic nonlinearity is primarily induced by the sigmoid-like high-Q power transmission response illustrated in Fig. 4(a) as the zoom-in version of Fig. 3(a) for a-MRM$_{11}$ simulated by the MRM models in Verilog-A [35], [36] although the transmission resonance can be linearly shifted with the reverse bias driven by a linear DAC.

For instance, the wavelength of $\lambda_1$ light-wave, whose power carries y$_1$<3:0> information, is designed at 1534.5 nm to match the resonance wavelength of a-MRM$_{11}$ under a certain reverse bias generated by a linear 4-bit DAC when a$_{11}$<3:0> = <0000>. Then, the a-MRM$_{11}$ resonance wavelength and transmission response can be horizontally shifted in a constant rate of 0.04 nm/V when any one of the rest 15 possible reverse biases is generated by this linear 4-bit DAC according to a$_{11}$<3:0> varying between <0001> and <1111> as shown in Fig. 4(a). Unfortunately, this horizontally linear shift causes the power attenuation (i.e., power gain < 1) of the $\lambda_1$ light-wave nonlinearly distributed across the vertical E/O DR as marked by black triangles in Fig. 4(a) and 4(b) on a dB scale. The same a$_{11}$<3:0> vs. E/O conversion curves are also shown in Fig. 4(e) but on a linear scale for the sake of linearity demonstration. To alleviate this nonlinear power modulation on the $\lambda_1$ light-wave, an additional calibration bit is added to basically perform a 5-bit linear DAC and generate 32 reverse-bias options so that the encoder and calibration logic can pick 16 out of 32 reverse biases to linearize the E/O transfer curve as marked by the blue circles in Fig. 4(e) and 4(b). Equivalently, all 16 values of a$_{11}$<3:0> are nonlinearly mapped to 16 values of a'$_{11}$<3:-1> to inversely cancel the nonlinear power gain effect of the MRM and to eventually obtain better linearity in the E/O conversion. In other words, a$_{11}$<3:0> is mapped to 16 out of 32 of a'$_{11}$<3:-1> to shift the transmission response nonlinearly as shown in Fig. 4(d), so that the power attenuation of $\lambda_1$ light-wave based on the values of a$_{11}$<3:0> relatively has better linearity. The linearity improvement of this technique is quantified by the deterministic E/O differential nonlinearity (DNL) and integral nonlinearity (INL) in Fig. 4(c) and 4(f), respectively, which are both reduced down to within ±LSB/2 of the 4-bit E/O conversion process.

### E. Racetrack-Resonator Photodetectors & O/E Conversions

The monolithic SiPh fabrication technology brings tremendous improvements for integration and energy efficiency of electronic-photonic circuits and systems. However, the major downsides are the degradations of photonic characteristics due to zero change to the underlying CMOS process. For instance, the responsivity of a 50-μm×5-μm linear PD is only about 0.023 A/W in 45SPCLO [14]. A solution utilizing the property of resonance amplification has been proposed to boost the responsivity of a micro-ring-resonator (MRR) based photodetection up to 0.55 A/W [7], [37] in this process. Unfortunately, because of the single wavelength selectivity of a MRR, this solution is only helpful to the WDM-based communication requiring to distinguish the powers of individual wavelengths as discussed in Section IV.A.

To boost the PD efficiency in 45SPCLO and simultaneously detect the aggregate light-wave power of all wavelengths used in the WDM-based computation for the proposed M-SiPh MVM accelerator, a broadband racetrack-resonator (RTR) [38] based PD is adopted as shown in Fig. 1, 2, and 3 [49]. The concepts of RTRs and MRRs are basically identical, but the dimensions of their resonance cavity and corresponding FSRs are quite different. As shown in Fig. 3(b), if the waveguide widths and gaps of the MRR and RTR are identical, their resonance-cavity lengths, $L_j$ and $L_{RTR,i}$, are the key design parameters for the power transmission and absorption responses.

This paragraph lists three examples to clarify the difference between the power transmission and absorption based on the concept of resonance effect since both play the key roles in the E/O and O/E conversions of the M-SiPh accelerators. As shown in the top-left of Fig. 3(b) for the 1st case, a certain amount of light-wave power of a certain wavelength from the Input port of an MRR is coupled into the resonance cavity (i.e., ring waveguide) of the MRR according to the reverse bias of the MRR P/N junction. Ideally, the light-wave power coupled into the resonance cavity can get amplified due to the resonance effect, and the residual light-wave power is transmitted to the Through port of this MRR. The MRR in this case is essentially configured as an MRM, and the reverse bias controls the refractive index $n_{eff}$ of the ring waveguide to shift the wavelength of the critical coupling as discussed in Section IV.D. The power transmission response is the defined by the ratio of the Through-port to Input-port powers. The light-wave power resonating in the MRR resonance cavity can be utilized in two different ways. As shown in the top-middle of Fig. 3(b) for the 2nd case, if another straight waveguide offers the Drop port for the MRR, the light-wave power in the resonance cavity can be further coupled to the Drop port and transmitted. The MRM with a Drop-port configuration is useful to the WDM-based communication discussed in Section IV.A and M-SiPh MMA elaborated in Section VI requiring to extract the information carried by a certain wavelength. In this case, the power transmission response is the defined by the ratio of the Drop-port to Input-port powers. As shown in the top-right of Fig. 3(b) for the 3rd case, instead of adding a Drop port, a SiGe layer (highlighted in olive green) is added between the ring waveguide and segmented P/N junction so that the amplified light-wave power in the resonance cavity can be effectively absorbed by the P/N junction to stimulate electron-hole pairs for generating a photocurrent, which is similar to a linear PD but exploits the power amplification of the resonance effect to boost the equivalent photodetection responsivity [37]. In this case, the power absorption response is the defined by the ratio of the absorbed to Input-port powers. Note that, in the 3rd case, the reverse bias of the P/N junction simultaneously affects the MRR critical coupling condition and PD responsivity. However, a single bias usually cannot meet the criteria for both, so the thermal control mechanism discussed in Section IV.H is necessarily required to provide an additional control knob for adjusting the critical coupling condition when the reverse bias

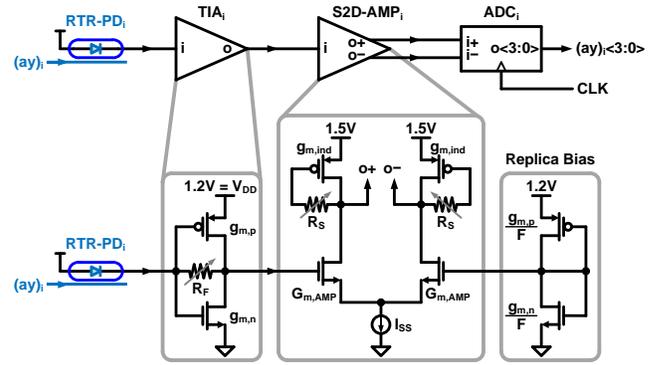

Fig. 5. The block diagram of each M-SiPh MVM dot-product O/E conversion circuit, and the schematics of $TIA_i$ and $S2D\text{-}AMP_i$.

is used for maximizing the responsivity to reach overall optimized power absorption.

For the M-SiPh MVM accelerator, an RTR-PD has to not only boost its O/E responsivity but also absorb all light-wave powers across the whole WDM spectrum as shown in the bottoms of Fig. 3(a) and 3(b). In other words, the $RTR\text{-}PD_i$ resonance cavity simultaneously establishes constructive inferences with all $\lambda_j$ wavelengths so that each MVM dot-product result carried by all $\lambda_j$ light-wave powers in each MVM row can be altogether absorbed by the P/N junction underneath the racetrack waveguide thorough the SiGe layer. On the condition of critical coupling across all $\lambda_j$, the aggregate optical power received by $RTR\text{-}PD_i$ is the linear summation of all $\lambda_j$ power amplified by the concurrent multi-wavelength resonances in the common racetrack cavity, which is the key idea of enhancing O/E responsivity for the broadband photodetection as follows:

$$P_{Absorb\text{-}RTR,i} \approx \frac{Q_{RTR,i}}{\pi \cdot L_{RTR,i}} \cdot \sum_{j=1}^{M} \frac{\lambda_j \cdot P_{trans\text{-}MRM,ij}(a_{ij})}{n_g(\lambda_j)} \quad (9)$$

To reach a consistent power absorption gain for all WDM wavelengths as shown in the bottom of Fig. 3(a), the perimeter $L_{RTR,i}$ of $RTR\text{-}PD_i$ for the whole WDM spectrum needs to meet two fundamental requirements: (I) The FSR $\Delta\lambda_{FSR\text{-}RTRj}$ of each resonance wavelength $\lambda_j$ determines each WDM isolation spacing $\Delta\lambda_{WDM,j}$ [33]. (II) Every resonance wavelength $\lambda_j$ under a particular resonance mode $m_{RTR,j}$ (an integer) is based on its effective refractive index $n_{eff}(\lambda_j)$, silicon propagation constant $\beta(\lambda_j)$, and the common perimeter $L_{RTR,i} = 2\pi \cdot r_{RTR,i} + 2 \cdot s_{RTR,i}$ [34]. These two requirements are expressed in Eq. (10) and (11) as follows:

$$\Delta\lambda_{FSR\text{-}RTR,j} = \frac{\lambda_j^2}{n_g(\lambda_j) \cdot L_{RTR,j}} = \Delta\lambda_{WDM,j}$$

$$\Rightarrow L_{RTR,i} = \frac{\lambda_j^2}{n_g(\lambda_j) \cdot \Delta\lambda_{WDM,j}} \quad (10)$$

$$2\pi \cdot m_{RTR,j} = \beta(\lambda_j) \cdot L_{RTR,i} = \frac{2\pi}{\lambda_j} \cdot n_{eff}(\lambda_j) \cdot L_{RTR,i}$$

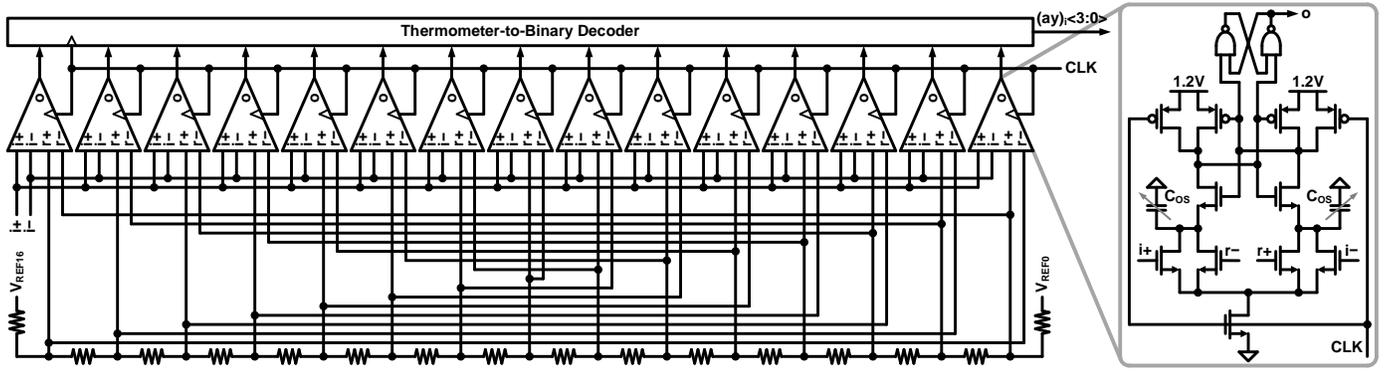

Fig. 6. The schematic of ADC$_i$ in each M-SiPh MVM dot-product O/E conversion circuit, and the schematic of the SAL-based clocked comparator.

$$\Rightarrow L_{RTR,i} = \frac{\lambda_j}{n_{eff}(\lambda_j)} \cdot m_{RTR,j} \quad (11)$$

After combining Eq. (10) and (11), all resonance wavelengths $\lambda_j$, $j = 1 \ldots M$, of the WDM spectrum are determined by the resonance mode requirements of RTR-PD$_i$ as follows:

$$m_{RTR,j} = \frac{n_{eff}(\lambda_j)}{n_g(\lambda_j)} \cdot \frac{\lambda_j}{\Delta\lambda_{WDM,j}}$$
$$= \frac{n_{eff}(\lambda_j)}{n_{eff}(\lambda_j) - \lambda_j \cdot \frac{\partial n_{eff}(\lambda_j)}{\partial \lambda_j}} \cdot \frac{\lambda_j}{\Delta\lambda_{WDM,j}} \quad (12)$$

Because all constructive interferences simultaneously occur in the same resonance cavity $L_{RTR,i}$, the only way to distinguish "$M$" resonance wavelengths $\lambda_j$ by Eq. (11) with a common $L_{RTR,i}$ is done by assigning "$M$" individual $m_{RTR,j}$, $j = 1 \ldots M$. In sum, the value of "$M$" and the requirements of Eq. (10), (11), and (12) determine the distribution of $\lambda_j$ and $\Delta\lambda_{WDM,j}$ of the WDM wavelength spectrum for the MVM operation; then the laser wavelengths and the radii of all y-MRM$_i$ and a-MRM$_{ij}$ described in Section IV.D shall be designed correspondingly to match the power transmission and absorption spectrums as illustrated in Fig. 3(a).

As shown in Fig. 3(a) and Table I, the consecutive integer $m_{RTR,j}$, $j = 1 \ldots 32$, are chosen from 2321 down to 2290 so that $\lambda_j$ and $\Delta\lambda_{WDM,j}$, $j = 1 \ldots 32$, can reach the targeted 0.5-nm WDM isolation spacing with $L_{RTR,i} = 951.32$ μm. About the footprint of RTR-PD$_i$ shown in the bottom of Fig. 3(b), if the radius $r_{RTR,i}$ of the left/right-end half-circles is set to 5 μm, the length $s_{RTR,i}$ of the top/bottom straight waveguides is 459.95 μm. Overall, including the primary racetrack-resonator and peripheral keep-out halo, the silicon area of RTR-PD$_i$ is within a 480-μm×20-μm tile.

*F. Transimpedance Amplifiers, Single-Ended-to-Differential Amplifiers & Analog-to-Digital Data Converters*

After the dot-product (ay)$_i$ converted from the aggregate light-wave power to a photocurrent in the electronic domain through RTR-PD$_i$, the following TIA$_i$, single-end-to-differential amplifier (S2D-AMP$_i$), and ADC$_i$ as illustrated in Fig. 5, further convert the format of (ay)$_i$ from the photocurrent to a L-bit digital word, where the entire MVM operation is basically completed.

The 1$^{st}$ stage, TIA$_i$, in Fig. 5 is a voltage-to-current feedback-amplifier architecture, whose feedforward path is a complimentary P/N transconductance ($G_{m,TIA} = g_{m,p} + g_{m,n}$) stage formed by a pair of P/N transistors conducting a self-biased DC current from V$_{DD}$ to GND, and the feedback path is a resistor R$_F$ playing the key role of the TIA gain, bandwidth, and input/output impedance with $G_{m,TIA}$. Because of the high-bandwidth and low-power characteristics, this TIA architecture has been widely used in high-speed optical receiver front-ends [5], [39]. Its simplified output resistance R$_{TIA}$, transfer function TF$_{TIA}(s)$, and average input-referred noise power spectrum density (PSD) $\overline{I^2_{n,in,TIA}}$ [39] are summarized as follows:

$$R_{TIA} \approx \frac{R_F}{1 + G_{m,TIA} \cdot R_F} \quad (13)$$

$$TF_{TIA}(s) \approx -\frac{G_{m,TIA} \cdot R_F \cdot R_{TIA}}{1 + s \cdot R_{TIA} \cdot C_{TIA}} \quad (14)$$

$$\overline{I^2_{n,in,TIA}} \approx \frac{2\pi \cdot \kappa \cdot T}{R_F^2} \cdot \left(\frac{\gamma}{G_{m,TIA}} + \frac{1}{G^2_{m,TIA} \cdot R_{TIA}}\right) \quad (15)$$

where Boltzmann constant $\kappa = 1.38 \times 10^{-23}$ J/K; thermal dynamic temperature T = 300 K; excess noise coefficient for deep submicron technology $\gamma \approx 2.5$; capacitive load at the TIA$_i$ output $C_{TIA} \approx 30$ fF, which is mainly the input capacitance of S2D-AMP$_i$. If the RTR-PD$_i$ input DR and O/E responsivity are 670-μW and 0.5-A/W, respectively, then the TIA input DR needs to be 335 μA. Meanwhile, if the TIA$_i$ output DR is set to 335 mV, $G_{m,TIA}$ and R$_F$ are chosen to be 1 mA/V and 1.65 kΩ, respectively, so that the TIA DC gain can be about 335-mV/335-μA = 1 kΩ ≈ |TF$_{TIA}(0)$| according to Eq. (14). These design specifications also lead to the TIA$_i$ bandwidth ≈ $1/(2\pi \cdot R_{TIA} \cdot C_{TIA})$ = 8.52 GHz, which is sufficient for the 1-GHz Nyquist frequency of the 2-GSym/s per (ay)$_i$ data. More importantly, the average input-referred noise PSD of TIA$_i$ as the first electronic circuit stage is less than 6.26 pA/√Hz based on Eq. (15). About the average power consumption, TIA$_i$ and its scaled replica consume about 0.1 mW from the 1.2-V supply to support the required 1-mV/A $G_{m,TIA}$ and DC common-mode voltages for the S2D-AMP$_i$ input differential pair.

The 2$^{nd}$ stage, S2D-AMP$_i$, in Fig. 5 is formed by a common-source differential amplifier with a pair of active-inductor loads to convert and buffer the single-ended TIA output to a differential signal for minimizing asymmetrical kickbacks and common-mode noise at the input of the following $ADC_i$. The active-inductor load [40] is a diode-connected PMOS with a tunable Drain-to-Gate feedback resistor $R_S$ to boost the high-frequency gain for linearly equalizing the data symbol and offer unbalanced gains to the positive and negative outputs for cancelling the nonideal single-ended-to-differential conversion process caused by the finite output impedance of the tail current source $I_{SS}$. If the S2D-$AMP_i$ output capacitance $C_{AMP}$, mainly contributed by the input of $ADC_i$, is 80 fF, and the S2D-$AMP_i$ output resistance $R_{AMP}$ is chosen to be 600 Ω, then the S2D-$AMP_i$ output bandwidth ≈ $1/(2\pi \cdot R_{AMP} \cdot C_{AMP})$ = 3.32 GHz, which could be the speed bottleneck of the entire MVM operation. Fortunately, the active-inductor load can effectively extend the bandwidth up to 5.5 GHz without extra power consumption, which is sufficient for the 1-GHz Nyquist frequency of the 2-GSym/s per $(ay)_i$ data. To convert the 335-mV $TIA_i$ output DR to 1-$V_{diff}$ $ADC_i$ input DR through S2D-$AMP_i$, the DC voltage-gain $|TF_{AMP}(0)|$ = $G_{m,AMP} \cdot R_{AMP}$ of S2D-$AMP_i$ is designated to 3×, so the transconductance $G_{m,AMP}$ and the static power consumption of S2D-$AMP_i$ can be determined as 5 mA/V and 0.75 mW, respectively.

The 3rd stage, $ADC_i$, in Fig. 5 is implemented by a L-bit flash ADC architecture containing $(2^L - 1)$ strong-arm-latch (SAL) based clocked-comparators [41] as shown in Fig. 6, which is suitable for high-speed and low-resolution applications with the downside of $2^L$ exponential-growth of the input capacitance, circuit area, and dynamic power. In the case of this 4-bit M-SiPh MVM accelerator, the 15 comparators in $ADC_i$ actually reach adequate compromise between area and 2-GS/s conversion rate with negligible static power. Each SAL-based clocked comparator contains a SAL and RS-latch to form an edge sampled DFF. The dual differential-pairs of each SAL are used to compare the analog volage difference between the $ADC_i$ differential-input and differential-reference voltages generated by a global resistor ladder for all ADCs. The SAL itself is capable of completing signal integral, regeneration, and decision within a half-period of the sampling clock with a single-phase sampling clock so the $ADC_i$ average power consumption is about 15×80-μW = 1.2 mW. Also, within each SAL, 3-bit capacitor-banks $C_{OS}$ are required for individual offset cancellations.

The low-power O/E circuit design criteria are bounded by the A/D accuracy along with the specification of the laser injection power for the WDM-based computation. Therefore, the design procedure shall consider the major noise contributors from RTR-$PD_i$ shot noise ($\overline{I_{n,PD}^2}$), $TIA_i$ circuit output noise ($\overline{V_{n,TIA}^2}$), and S2D-$AMP_i$ circuit output noise ($\overline{V_{n,AMP}^2}$). The overall noise power at the $ADC_i$ input can be approximated as follows:

$$P_{n,O/E} \approx \int_0^\infty \overline{I_{n,PD}^2}(f) \cdot |TF_{TIA}(f)|^2 \cdot |TF_{AMP}(f)|^2 \cdot df$$

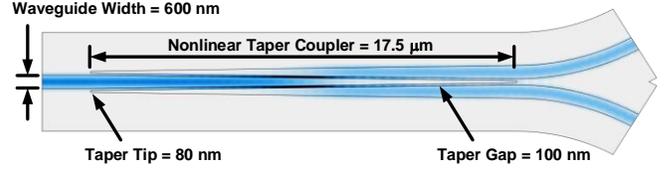

Fig. 7. The mask layout and dimensions of the 50/50 Y-junction power-splitter unit with the nonlinear taper coupling technique.

$$+ \int_0^\infty \overline{V_{n,TIA}^2}(f) \cdot |TF_{AMP}(f)|^2 \cdot df + \int_0^\infty \overline{V_{n,AMP}^2}(f) \cdot df$$
$$\approx 0.5qI_{PD}(G_{m,TIA}^2 R_F^2 R_{TIA}^2)G_{m,AMP}^2 R_{AMP}/C_{AMP}$$
$$+ kT(\gamma G_{m,TIA}R_{TIA}^2 + R_{TIA})G_{m,AMP}^2 R_{AMP}/C_{AMP}$$
$$+ 2kT(\gamma G_{m,AMP}R_{AMP} + \gamma g_{m,ind}R_{AMP} + 1)/C_{AMP} \quad (16)$$

where the elementary charge q = $1.602 \times 10^{-19}$ C; $g_{m,ind}$ is the transconductance of the PMOS for the active-inductor; the S2D-$AMP_i$ output resistance $R_{AMP} \approx R_S/(1+g_{m,ind} \cdot R_S) \approx 600$ Ω. The three frequency-domain integrals in Eq. (16) are simplified by only considering the S2D-$AMP_i$ bandwidth due to its speed domination. If the maximum $I_{PD} \approx 335$ μA = the TIA input DR as mentioned, the total noise power at the $ADC_i$ input estimated by Eq. (16) is about 11 μW. Compared to the 4-bit quantization noise power of $ADC_i$ = $V_{LSB}^2/12$ = $(1-V_{diff}/15)^2/12$ = 370 μW, there is a margin of 2.5 bits ≈ 15.3 dB = $10 \cdot \log_{10}(370$-μW$/11$-μW$)$ to accommodate dark noise, flicker noise, supply noise, clock jitter, resistor-ladder noise, comparator noise, and residual offset excluded in the noise estimation of Eq. (16). Overall, the silicon area of $TIA_i$, S2D-$AMP_i$, and $ADC_i$ is about 100-μm×20-μm for each MVM row.

## G. Optical Power Splitters & Laser-Comb Injection Power

The M-SiPh MVM accelerator requires a laser comb source [20] to offer the WDM wavelengths $\lambda_i$, i = 1 … M as shown in Fig. 1. After the E/O conversion of the input vector $Y_{M \times 1}$, the 1-to-M O-PS evenly splits the pre-summed vector elements ($\sum_{i=1}^{M} y_i$) into "M" identical waveguides ready for the following WDM-based MVM operation. This 1-to-M O-PS contains $\log_2(M)$ horizontal stages, and each stage is vertically formed by multiple 50/50 PS units in parallel with a number from $2^0$ (= 1) to $2^{\log_2(M)-1}$ (= M/2) for the first to last stages, respectively. Each 50/50 PS unit is basically an adiabatic Y-junction with nonlinear taper coupling technique [42] for the characteristics of low-loss, high-bandwidth, high-polarization insensitivity, and high-tolerance to fabrication errors. The simulation result of a single 50/50 PS unit in Fig. 7 on the 160-nm silicon-on-insulator (SOI) platform in 45SPCLO shows that a 17.5-μm nonlinear taper coupler within a total 35-μm footprint, including the length from the single horizontal fan-in to two horizontal fan-outs, can reach about 0.07-dB [43] transverse electric (TE) transmission loss in the range from 1530-nm to 1550-nm WDM wavelength spectrum. The overall silicon area of the 1-to-M O-PS is $[\log_2(M) \cdot 35$-μm$] \times [M \cdot 20$-μm$]$, which is 175-μm×640-μm when M = 32.

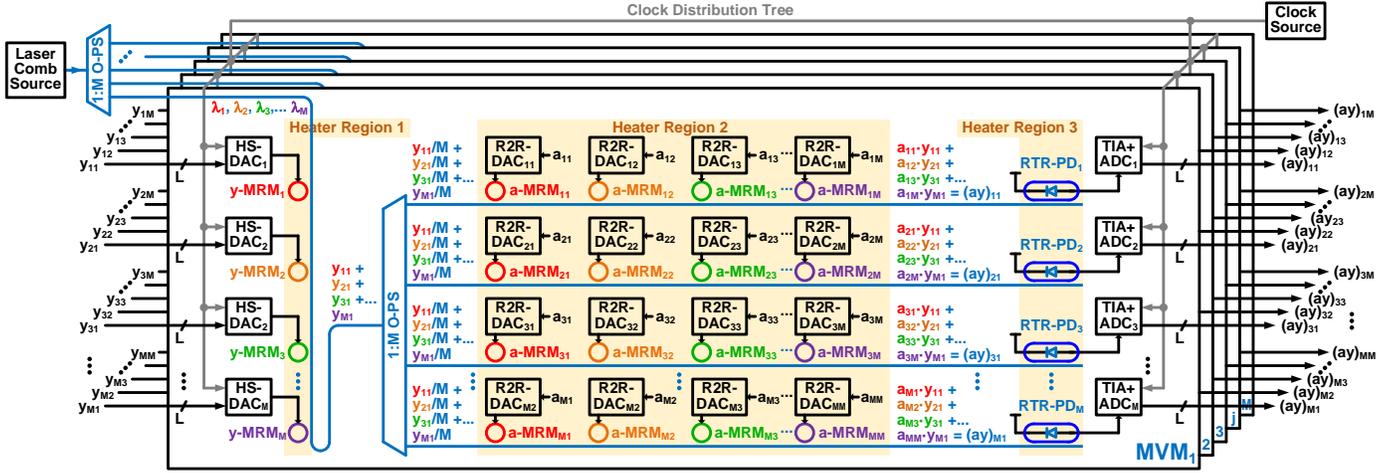

Fig. 8. The system block diagram of the M-SiPh MMM accelerator implemented by the fully M-SiPh MVM parallelism approach.

The aggregate laser power loss, $\log_2(M)\cdot(3.01 + 0.07)$-dB, of this 1-to-$M$ O-PS is consolidated with the transmission and absorption losses of MRMs and RTR-PD in each MVM row to estimate the required laser-injection power of each wavelength for the input DR and signal-to-noise ratio of O/E conversion. All laser wavelengths $\lambda_i$, $i = 1 \ldots M$, shall have identical injection power $P_\lambda$ and must propagate through $\log_2(M)$ horizontal stages of 50/50 Y-junctions (3.01-dB power splitting and 0.07-dB loss per Y-junction), two MRM transmission DR losses (2.5-dB loss per y-MRM$_i$ or a-MRM$_{ij}$), and one RTR-PD absorption DR loss (2.5-dB loss per RTR-PD$_i$). Meanwhile, the aggregate absorption of all wavelength powers per RTR-PD needs to be confined within the linear DR of the O/E conversion circuit, $DR_{O/E}$. In sum, the laser-injection power per wavelength can be expressed as follows:

$$P_\lambda \cdot \left(10^{\frac{-2.5\text{-}dB}{10}}\right)^3 \cdot \left(\frac{1}{M} \cdot 10^{\frac{-\log_2(M)\cdot 0.07\text{-}dB}{10}}\right) \cdot M \leq DR_{O/E} \quad (17)$$

Note that $P_\lambda$ is not a strong function of "$M$" since $DR_{O/E}$ stays constant regardless of "$M$", and only the number of the horizontal stages gradually accumulate the power loss. If $M = 32$ and $DR_{O/E} = 670$ μW, $P_\lambda$ is about 4.08 mW, and total 32 laser-injection power for this WDM-based MVM operation is 32×4.08-mW = 130.6 mW, which however is directly proportional to the number of WDM wavelengths "$M$".

*H. Thermal Tuning & Post-Fabrication Trimming*

Due to the high thermo-optic coefficient, $1.86\times10^{-4}$ 1/K, of silicon [44], proper thermo-control mechanisms are necessary to adjust the on-chip temperature for maintaining consistent characteristics of SiPh devices. Meanwhile, this high temperature sensitivity also enables high-resolution SiPh characteristic tunability beyond the fabrication resolution. For instance, in Table I, the 32 different radii of all a-MRM$_{ij}$ distributed from 4.63 μm to 4.76 μm cannot be explicitly fabricated by relying on the mask resolution of 45SPCLO. In CMOS compatible SiPh process technology, tungsten heaters have been widely used for thermo-optic tuning [45], and their power efficiency can reach about 2.4 mW/$\Delta\lambda_{FSR}$ for a single MRM [46]. Therefore, if an M-SiPh MVM accelerator consisting of "$M\cdot(1 + M)$" MRMs and "$M$" RTR-PDs can be fabricated within a 1-mm$^2$ silicon area, and the tuning ranges of each MRM and RTR-PD need to cover about one $\Delta\lambda_{WDM} \approx \Delta\lambda_{FSR}/M \approx \Delta\lambda_{FSR,RTR}$, the total tungsten heater power per M-SiPh MVM accelerator can be expressed as $M\cdot(1 + M)\cdot(2.4\text{-mW})/M + M\cdot2.4$-mW, which is 156 mW when $M = 32$.

To do the thermal control across a large number of MRMs in a high-dimensional M-SiPh MVM accelerator, one heater source per MRM is impractical, a hybrid tuning approach is proposed in [49] by combining the tungsten-heater approach for global coarse tuning with the post-fabrication-trimming approach [47] for individual fine tuning. For instance, in Fig. 1 and Fig. 8, all MRMs and RTR-PDs in the entire M-SiPh MVM accelerator are partitioned into three tungsten-heater regions with their own heater sources, and the area of each region shall be less than 1 mm$^2$ to limit the random variation among MRMs or RTR-PD less than one standard deviation within each region. In this way, a certain heater source of a certain heater region can globally control and shift the transmission/absorption responses of the MRMs/RTR-PDs. Then, the post-fabrication trimming mechanism can take care of low-range but high-accuracy fine tuning for every MRM and RTR-PD, which is basically realized by implanting a section of SOI rib waveguide with Ge through a photoresist mask on the top of each resonance cavity, so each MRM or RTR-PD critical coupling condition can be trimmed by injecting a voltage pulse to anneal this Ge rib waveguide. This annealing calibration process can be done for each MRM or RTR-PD individually as shown in Fig. 2, and the annealing pulse width for the targeted resonance wavelength(s) of each MRM or RTR-PD can be obtained by an iterative feedback mechanism of the MRM/RTR-PD transmission/absorption power received by its ADC to adjust the annealing pulse generator output pulse width until converging to the critical coupling condition [47]. Though this post-fabrication trimming approach is tedious, it is thorough, reprogrammable, hardware reusable, and it only consumes calibration time/power with negligible overhead during the regular M-SiPh accelerator operations.

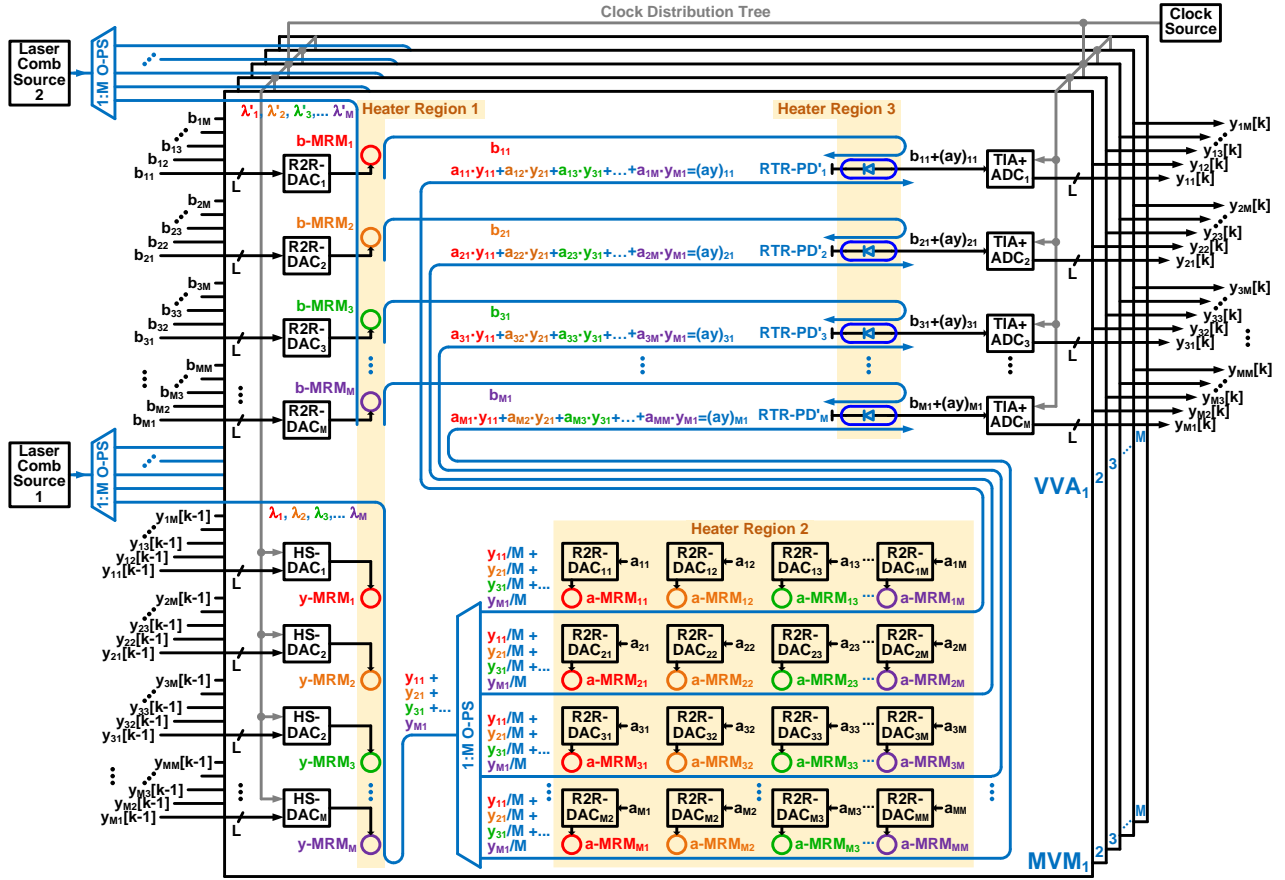

Fig. 9. The system block diagram of the M-SiPh MI accelerator implemented by the M-SiPh MVM and VVA parallelism approach.

## V. M-SiPh Matrix-Matrix Multiplications

To perform an M-SiPh MMM operation, $A_{M \times M} \cdot Y_{M \times M} = (AY)_{M \times M}$, the input $M$-by-$M$ matrix $Y_{M \times M}$ with elements denoted by $y_{ij}$, i and j = 1 … $M$ respectively, can be split into "$M$" $M$-by-1 column vectors as shown in Fig. 8, and each column vector independently performs MVM with matrix $A_{M \times M}$ in an M-SiPh MVM unit to produce its own output column vector. After combining total "$M$" output column vectors from "$M$" parallel M-SiPh MVMs, the outcome $M$-by-$M$ matrix $(AY)_{M \times M}$ with elements denoted by $(ay)_{ij}$, i and j = 1 … $M$ respectively, of the M-SiPh MMM operation can be obtained.

This hardware parallelism approach can be equivalently implemented in a manner of time-multiplexing. For instance, a single M-SiPh MVM unit can process one of the input column vectors per iteration period, and electronic registers after the ADCs can collect all the output column vectors after "$M$" iteration cycles to form the final outcome matrix $(AY)_{M \times M}$ per M-SiPh MMM operation. These two formats of the M-SiPh MMM implementation linearly consume the silicon area and computation time with the column-dimension "$M$" of $Y_{M \times M}$, respectively. In sum, energy/area/throughput tradeoffs can be optimized by combining the partial hardware-parallelism and partial time-multiplexing approaches based on realistic applications.

## VI. M-SiPh Matrix-Inversions

Each repetition of the Neumann-series approximation, $Y_{M \times M}[k] = B_{M \times M} + A_{M \times M} \cdot Y_{M \times M}[k-1]$, for the targeted matrix-inversion (MI) functionality can be energy/area efficiently implemented by the combination of M-SiPh MMM and M-SiPh MMA without intermediate O/E/O conversions as shown in Fig. 9, where the matrices $A_{M \times M}$, $B_{M \times M}$, $Y_{M \times M}[k]$, and $Y_{M \times M}[k-1]$ are physically presented by their matrix elements $a_{ij}$, $b_{ij}$, $y_{ij}[k]$, and $y_{ij}[k-1]$, i and j = 1 … $M$ respectively. The bottom-half of Fig. 9 illustrates the MMM operation of $A_{M \times M} \cdot Y_{M \times M}[k-1]$, which is essentially realized by the M-SiPh MMM accelerator elaborated in Section V however without the O/E conversion to keep the MMM results in the photonic domain. Then, the top-half of Fig. 9 shows the MMA operation for the completion of $Y_{M \times M}[k]$ achieved by element-to-element light-wave power additions of $b_{ij}$ and $(ay)_{ij}$, i and j = 1 … $M$, respectively.

Similar to the parallelism architecture for realizing an MMM from multiple MVM units in Section V, the M-SiPh MMA can be implemented by "$M$" parallel vector-vector additions (VVA) as shown in the top-half of Fig. 9. The 2nd laser comb source is required to generate another set of "$M$" wavelengths $\lambda'_i$, i = 1 … $M$, whose WDM spectrum relationship with $\lambda_i$, i = 1 … $M$, from the 1st laser comb source is reflected in Fig. 10(a). Accordingly, the power transmission responses of b-MRM$_i$, i = 1 … $M$, have $(\Delta\lambda_{WDM,i})/2$ offsets with respect to those of y-

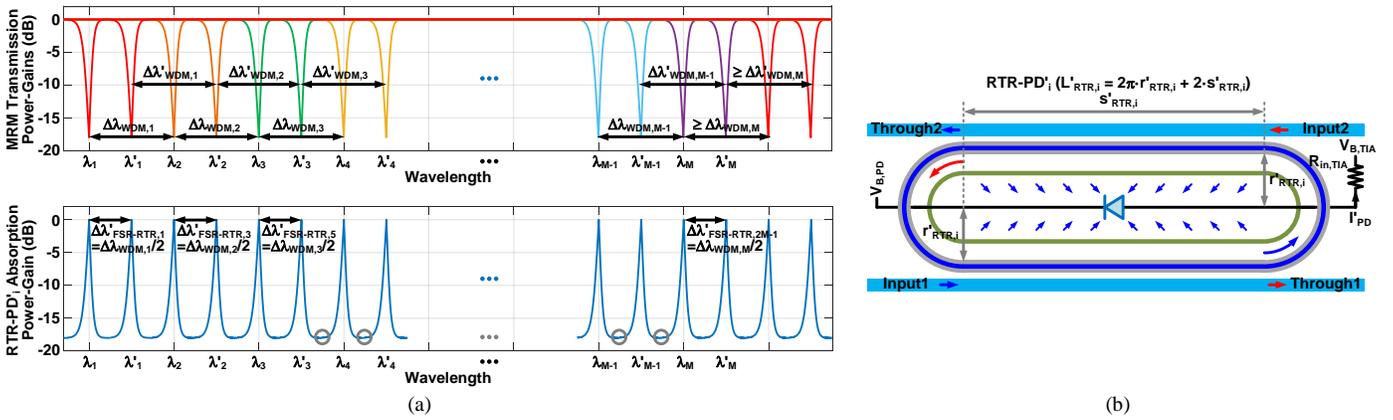

Fig. 10. (a) The transmission responses of y-MRM$_i$ and b-MRM$_i$, i = 1 … M, and the absorption responses of RTR-PD'$_i$. (b) The conceptual geometry and configuration of RTR-PD'$_i$.

TABLE II
PERFORMANCE ESTIMATIONS OF M-SIPH MVM VS. ASIC MVM

| | Vector Dimen. "M" | Laser Power (mW) | Heater Power (mW) | SoC Power* (mW) | SoC Area (mm$^2$) | Data Precision (bit) | Clock Rate (GHz) | Computation Throughput | | Computation Density (TMAC/s/mm$^2$) | Energy Consumption (fJ/MAC)* |
|---|---|---|---|---|---|---|---|---|---|---|---|
| | | | | | | | | TOPS/s | TMAC/s | | |
| **M-SiPh MVM** in 45-nm Monolithic SiPh [This Work] | 8 | 31.6 | 40.8 | 99.6 | 0.10 | 4 | 2 | 0.256 | 0.128 | 1.26 | 777.8 |
| | 16 | 64.3 | 79.2 | 198.7 | 0.33 | 4 | 2 | 1.024 | 0.512 | 1.56 | 388.0 |
| | 32 | 130.7 | 156.0 | 400.7 | 1.14 | 4 | 2 | 4.096 | 2.048 | 1.80 | 195.6 |
| | 64 | 265.6 | 309.6 | 818.0 | 4.16 | 4 | 2 | 16.384 | 8.192 | 1.97 | 99.8 |
| | 128 | 539.9 | 616.8 | 1701.1 | 15.77 | 4 | 2 | 65.536 | 32.768 | 2.08 | 51.9 |
| | 256 | 1097.3 | 1231.2 | 3653.3 | 61.12 | 4 | 2 | 262.144 | 131.072 | 2.14 | 27.9 |
| **ASIC MVM** Google TPUv4 in 7-nm CMOS [48] | Vector Dimen. "M" | SoC Idle Power (mW) | | SoC Busy Power (mW)† | SoC Area (mm$^2$) | Data Precision (bit) | Clock Rate (GHz) | Computation Throughput | | Computation Density (TMAC/s/mm$^2$) | Energy Consumption (fJ/MAC)† |
| | | | | | | | | TOPS/s | TMAC/s | | |
| | 256 | 55000 | | 78571‡ | 400 | 8 | 1.05 | 137.62 | 68.81 | 0.17 | 1141.9‡ |

\* Including all photonic/electronic devices/circuits power consumption, heater power, and laser injection power on a single M-SiPh chip.
† Including all electronic digital circuits power consumption on the single CMOS chip in the Busy-mode.
‡ The Busy-mode power of TPUv4 is estimated by its Idle-mode power (= 55 W) and Busy-vs.-Idle power ratio of TPUv1 (= 1.43) [49].

MRM$_i$, i = 1 … M, to enable incoherent detections in RTR-PD'$_i$. Note that each b-MRM can be driven by a low-power R2R-DAC due to the static value of b$_{ij}$ in the electronic domain. Also, after the E/O conversion through b-MRM$_i$, the light-wave λ'$_i$ carrying the b$_{ij}$ information in the photonic domain is individually extracted from the rest of WDM wavelengths at the Drop port of b-MRM$_i$ as the MRR configuration shown in the top-middle of the Fig. 3(b) for the following M-SiPh VVA operation.

The power absorption spectrum of RTR-PD'$_i$ shall cover and align with all resonance wavelengths λ$_i$ and λ'$_i$ of y-MRM$_i$ and b-MRM$_i$, i = 1 … M, as shown in the bottom of Fig. 10(a) to incoherently absorb all different light-wave powers carrying the (ay)$_{ij}$ and b$_{ij}$ information. Equivalently, to perform Δλ'$_{FSR\text{-}RTR,2i\text{-}1,2i}$ = (Δλ$_{WDM,i}$)/2, i = 1 … M, the resonance-cavity length of RTR-PD'$_i$ has to be doubled compared to RTR-PD$_i$ discussed in Section V.E; i.e., L'$_{RTR,i}$ = 2×L$_{RTR,i}$. Meanwhile, reducing the FSR of RTR-PD'$_i$ by 2× could indeed cause overlaps between adjacent spectrum "skirts" of power absorption responses as circled in Fig. 10(a), but these don't create crosstalk among the wavelengths since the WDM spectrum isolations are still maintained by Δλ$_{WDM,i}$ and Δλ'$_{WDM,i}$, i = 1 … M, individually. Also, to couple both (ay)$_{ij}$ and b$_{ij}$ into the racetrack waveguide, RTR-PD'$_i$ requires two Input ports implemented by running two input waveguides in parallel with the top and bottom straight waveguides of RTR-PD'$_i$ as shown in Fig. 9 and 10(b), where the directions of the incoming light-waves carrying (ay)$_{ij}$ and b$_{ij}$ information on these two input waveguides should form consistent clockwise or counterclockwise coupling to the racetrack waveguide for minimizing crosstalk between these two Input ports due to imperfect absorption leaking through the Input-to-Drop responses of RTR-PD'$_i$.

Overall, the proposed M-SiPh MI accelerator can effectively accomplish the end-to-end computation of each Neumann-series repetition by executing computation-dominant operations in the photonic domain at the speed of light with a minimal amount of energy/area overhead due to the monolithic photonic-electronic on-chip integrations, conversions and calibrations.

## VII. PERFORMANCE SUMMARY AND CONCLUSION

The key performance metrices of high-performance computing systems include computation throughput (TMAC/s), computation density (TMAC/s/mm$^2$), and energy consumption (fJ/MAC) [9], [13], [18]. By consolidating the power, area, and clock rate of the proposed building blocks described and analyzed in Section IV, the detailed energy/area breakdowns and computing performance metrics of the M-SiPh MVM accelerator with the on-chip hardware and MVM functionality are summarized in Table II to practically cover the overhead of the M-SiPh integrations, conversions, and calibrations, including all electronic-photonic SoC building blocks with

This work has been submitted to the IEEE for possible publication. Copyright may be transferred without notice, after which this version may no longer be accessible.15 complete DACs/ADCs, on-chip digital interface, clock distribution, on-chip calibration hardware, laser injection power, and heater power. For the comparison purpose, the performance metrics of the Google Tensor Processing Units v4 (TPUv4) [48] are listed in Table II as well since this TPU is also a complete on-chip MVM accelerator but implemented by a digital ASIC approach in commercial 7-nm CMOS process technology. Note that the performance metrics of the M-SiPh MMM and M-SiPh MI accelerators can be reasonably estimated according to those of the M-SiPh MVM accelerator because of the dimension scalability and parallelism architecture described in Section V and VI.

The performance scalability with the input vector dimension "$M$" of the M-SiPh MVM accelerator listed in Table II shows that the energy/area overhead of M-SiPh MVM is getting leveraged by the negligible photonic computing latency and energy consumption when "$M$" is scaling up though inevitable increase of the calibration complexity is not reflected by these performance matrices in the regular accelerator operations. In the cases of "$M$" ≥ 8, the M-SiPh MVM accelerator outperforms the ASIC counterpart in both computation density and energy consumption. In particular with the future advances in scaling the optical comb generations to "$M$" = 256, the M-SiPh MVM accelerator can exhibit about 12.6× computation-density and 40.9× energy-efficiency superiority [49] over the advanced ASIC MVM accelerator (TPUv4) with the downside of lower data precision due to the overhead of "analog" computing in the photonic domain. Overall, given that novel SiPh devices and circuits are still being discovered and engineered for future foundry manufacturing, the performance of the M-SiPh accelerators will be further enhanced with the development of next-generation SiPh process technology.

## REFERENCES

[1] P. Moorhead, "Silicon Photonics are here and Global Foundries is innovating," Moor Insights & Strategy, Oct. 2022. [Online]. Available: https://gf.com/wp-content/uploads/2022/12/Silicon-Photonics-Are-Here-And-Global-Foundries-Is-Innovating-Final-V10.24.2022-2.pdf

[2] M. Rakowski, C. Meagher, K. Nummy, A. Aboketaf, J. Ayala, Y. Bian, B. Harris, K. Mclean, K. McStay, A. Sahin, L. Medina, B. Peng, Z. Sowinski, A. Stricker, T. Houghton, C. Hedges, K. Giewont, A. Jacob, T. Letavic, D. Riggs, A. Yu, and J. Pellerin, "45nm CMOS - Silicon Photonics Monolithic Technology (45CLO) for next-generation, low power and high speed optical interconnects," in *Optical Fiber Communication Conference (OFC)*, San Diego, CA, USA, 2020.

[3] C. Levy, Z. Xuan, D. Huang, R. Kumar, J. Sharma, T. Kim, C. Ma, G.-L. Su, S. Liu, J. Kim, X. Wu, G. Balamurugan, H. Rong, and J. Jaussi, "A 3D-integrated 8λ × 32 Gbps λ silicon photonic microring-based DWDM transmitter," in *IEEE Custom Integrated Circuits Conference (CICC)*, San Antonio, TX, USA, 2023.

[4] H. Li, G. Balamurugan, M. Sakib, J. Sun, J. Driscoll, R. Kumar, H. Jayatilleka, H. Rong, J. Jaussi, and B. Casper, "A 112 Gb/s PAM4 silicon photonics transmitter with microring modulator and CMOS driver," *IEEE Journal of Lightwave Technology*, vol. 38, no. 1, pp. 131–138, Jan. 2020.

[5] H. Li, G. Balamurugan, J. Jaussi, and B. Casper, "A 112 Gb/s PAM4 linear TIA with 0.96 pJ/bit energy efficiency in 28 nm CMOS," in *IEEE European Solid-State Circuits Conference (ESSCIRC)*, Dresden, Germany, 2018.

[6] C. Sun, M. Wade, M. Georgas, S. Lin, L. Alloatti, B. Moss, R. Kumar, A. Atabaki, F. Pavanello, J. Shainline, J. Orcutt, R. Ram, M. Popovic, and V. Stojanovic, "A 45 nm CMOS-SOI monolithic photonics platform with bit-statistics-based resonant microring thermal tuning," *IEEE Journal of Solid-State Circuits*, vol. 51, no. 4, pp. 893–907, April 2016.

[7] N. Mehta, S. Buchbinder, and V. Stojanović, "Design and characterization of monolithic microring resonator based photodetector in 45nm SOI CMOS," in *IEEE European Solid-State Device Research Conference (ESSDERC)*, Cracow, Poland, 2019.

[8] N. Mehta, C. Sun, M. Wade, S. Lin, M. Popović, and V. Stojanović, "A 12Gb/s, 8.6μApp input sensitivity, monolithic-integrated fully differential optical receiver in CMOS 45nm SOI process," in *IEEE European Solid-State Circuits Conference (ESSCIRC)*, Lausanne, Switzerland, 2016.

[9] M. A. Nahmias, T. F. de Lima, A. N. Tait, H.-T. Peng, B. J. Shastri, and P. R. Prucnal, "Photonic multiply-accumulate operations for neural networks," *IEEE J. Sel. Topics in Quantum Electronics*, vol. 26, no. 1, Jan.–Feb. 2020.

[10] K. Kikuchi, "Fundamentals of coherent optical fiber communications," *IEEE Journal of Lightwave Technology*, vol. 34, no. 1, pp. 157–179, Jan. 2016.

[11] C. Dragone, "Efficient N*N star couplers using Fourier optics," *IEEE Journal of Lightwave Technology*, vol. 7, no. 3, pp. 479–489 March 1989.

[12] R. A. Athale and W. C. Collins, "Optical matrix–matrix multiplier based on outer product decomposition," *Applied Optics*, vol. 21, no. 12, pp. 2089–2090, June 1982.

[13] H. Zhou, J. Dong, J. Cheng, W. Dong, C. Huang, Y. Shen, Q. Zhang, M. Gu, C. Qian, H. Chen, Z. Ruan, and X. Zhang, "Photonic matrix multiplication lights up photonic accelerator and beyond," *Light: Science & Applications*, vol. 11, no. 30, Feb. 2022.

[14] J. Feldmann, N. Youngblood, M. Karpov, H. Gehring, X. Li, M. Stappers, M. Le Gallo, X. Fu, A. Lukashchuk, A. S. Raja, J. Liu, C. D. Wright, A. Sebastian, T. J. Kippenberg, W. H. P. Pernice, and H. Bhaskaran, "Parallel convolutional processing using an integrated photonic tensor core," *Nature*, vol. 589, pp. 52–58, Jan. 2021.

[15] L. Yang, L. Zhang, and R. Ji, "On-chip optical matrix-vector multiplier for parallel computation," *International Society for Optics and Photonics*, June 2013, DOI: 10.1117/2.1201306.004932.

[16] D. Dang, B. Lin, and D. Sahoo, "LiteCON: An all-photonic neuromorphic accelerator for energy-efficient deep learning," *ACM Trans. Archit. Code Optim.*, vol. 19, no. 3, pp. 1–22, Aug. 2022.

[17] M. Li and Y. Wang. "An energy-efficient silicon photonic-assisted deep learning accelerator for big data." In *Proc. of the Conference on Wireless Communications and Mobile Computing*, 2020.

[18] C. Huang, V. J. Sorger, M. Miscuglio, M. Al-Qadasi, Avilash Mukherjee, L. Lampe, M. Nichols, A. N. Tait, T. F. de Lima, B. A. Marquez, J. Wang, L. Chrostowski, M. P. Fok, D. Brunner, S. Fang, S. Shekhar, P. R. Prucnal, and B. J. Shastri, "Prospects and applications of photonic neural networks." *Advances in Physics: X*, vol. 7, no. 1, pp. 1–63, 2022.

[19] B. J. Shastri, A. N. Tait, T. F. de Lima, W. H. P. Pernice, H. Bhaskaran, C. D. Wright, and P. R. Prucnal, "Photonics for artificial intelligence and neuromorphic computing." *Nature Photonics*, vol. 15, pp.102–114, Jan. 2021.

[20] H. Hu and L. K. Oxenløwe, "Chip-based optical frequency combs for high-capacity optical communications," *Nanophotonics*, vol. 10, no. 5, pp. 1367–1385, 2021.

[21] A. Burian, J. Takala, and M. Ylinen, "A fixed-point implementation of matrix inversion using Cholesky decomposition," in *Proc. of IEEE 46th Midwest Symposium on Circuits and Systems (MWSCAS)*, vol. 3, 2003, pp. 1431–1434.

[22] A. Rontogiannis, V. Kekatos, and K. Berberidis, "A square-root adaptive V-BLAST algorithm for fast time-varying MIMO channels," *IEEE Signal Processing Letters*, vol. 13, no. 5, pp. 265–268, 2006.

[23] M. Wu, B. Yin, A. Vosoughi, C. Studer, J. R. Cavallaro, and C. Dick, "Approximate matrix inversion for high-throughput data detection in the large-scale MIMO uplink," in *Proc. of the IEEE International Symposium on Circuits and Systems (ISCAS)*, 2013, pp. 2155–2158.

[24] X. Gao, L. Dai, Y. Hu, Z. Wang, and Z. Wang, "Matrix inversion-less signal detection using SOR method for uplink large-scale MIMO systems," in *Proc. IEEE Global Telecommun. Conf.*, Dec. 2014, pp. 3291–3295.

[25] L. Dai, X. Gao, X. Su, S. Han, C. L. I, and Z. Wang, "Low-complexity soft-output signal detection based on Gauss Seidel method for uplink multiuser large-scale MIMO systems," *IEEE Trans. Veh. Technol.*, vol. 64, no. 10, pp. 4839–4845, Oct. 2015.

[26] M. Wu, C. Dick, J. R. Cavallaro, and C. Studer, "FPGA design of a coordinate descent data detector for large-scale MU-MIMO," in *Proc. IEEE Int. Symp. on Circuits and Systems*, May 2016, pp. 1894–1897.

[27] B. Yin, M. Wu, J. R. Cavallaro, and C. Studer, "Conjugate gradient based soft-output detection and precoding in massive MIMO systems," in *Proc. IEEE Global Telecommun. Conf.*, Dec. 2014, pp. 3696–3701.

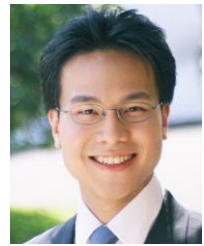

**Tzu-Chien Hsueh** (Senior Member, IEEE) received the Ph.D. degree in electrical and computer engineering from the University of California, Los Angeles, CA, in 2010. From 2001 to 2006, he was a Mixed-Signal Circuit Design Engineer in Hsinchu, Taiwan. From 2010 to 2018, he was a Research Scientist in Intel Lab Signaling Research and an Analog Engineer in Intel I/O Circuit Technology, Hillsboro, Oregon. Since 2018, he has been an Assistant Professor in electrical and computer engineering at the University of California, San Diego (UCSD). His research interests include wireline electrical/optical transceivers, clock-and-data recovery, data-conversion circuits, on-chip performance measurements/analyzers, and digital/mixed signal processing techniques.

Dr. Hsueh was a recipient of multiple Intel Division and Academy Awards from 2012 to 2018, the 2015 IEEE Journal of Solid-State Circuits (JSSC) Best Paper Award, the 2020 NSF CAREER Award, and the 2022 UCSD Best Teacher Award. He served on the Patent Committee for Intel Intellectual Property (Intel IP) and the Technical Committee for Intel Design & Test Technology Conference (DTTC) from 2016 to 2018. Since 2018, he has served on the Technical Program Committee for IEEE Custom Integrated Circuits Conference (CICC) and the Guest Associate Editor for IEEE Solid-State Circuits Letters (SSC-L).

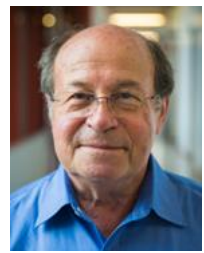

**Yeshaiahu (Shaya) Fainman** (Fellow, IEEE) is an inaugural ASML/Cymer Chair Professor of Advanced Optical Technologies and Distinguished Professor in Electrical and Computer Engineering (ECE) at the University of California, San Diego (UCSD). He received M. Sc and Ph. D degrees from Technion-Israel Institute of Technology in 1979 and 1983, respectively. He is directing research of the Ultrafast and Nanoscale Optics group at UCSD and made significant contributions to near field optical phenomena, nanoscale science and engineering of ultra-small, sub-micrometer semiconductor light emitters and nanolasers, inhomogeneous and meta-materials, nanophotonics and Si Photonics. His current research interests are in near field optical science and technology with Si Photonics applications to information technologies and biomedical sensing. He contributed over 340 manuscripts in peer review journals and over 560 conference presentations and conference proceedings. He contributed to editorial and conference committee works of various scientific societies including IEEE, SPIE and OPTICA. He is a Fellow of OPTICA (former OSA), Fellow of the IEEE, Fellow of the SPIE, and a recipient of the Miriam and Aharon Gutvirt Prize, Lady Davis Fellowship, Brown Award, Gabor Award, Emmett N. Leith Medal, Joseph Fraunhofer Award/Robert M. Burley Prize and OPTICA Holonyak Award.

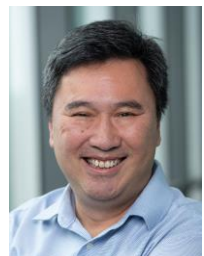

**Bill Lin** (Senior Member, IEEE) received the B.S., M.S., and Ph.D. degrees in electrical engineering and computer sciences from the University of California at Berkeley, Berkeley, CA, USA, in 1985, 1988, and 1991, respectively. He is currently a Professor in electrical and computer engineering with the University of California, San Diego, La Jolla, CA, USA, where he is actively involved with the Center for Wireless Communications (CWC), the Center for Networked Systems (CNS), and the Qualcomm Institute in industry-sponsored research efforts. Dr. Lin's research has led to more than 200 journal and conference publications, including a number of Best Paper awards and nominations. He also holds five awarded patents. He was the General Chair and on the executive and technical program committee of many IEEE and ACM conferences, and he was Associate and Guest Editors for several IEEE and ACM journals.